\documentclass[aps,prb,twocolumn,showpacs,amsmath,amssymb,floatfix]{revtex4}
\usepackage{color}
\usepackage{epsfig} \usepackage{bm}
\IfFileExists{srcltx.sty}{\usepackage[active]{srcltx}}

\begin{document}

\title{Mach-Zehnder interferometry of fractional quantum Hall edge states.}

\author{Ivan P. Levkivskyi$^{1,2}$, Alexey Boyarsky$^{3,4}$, J\"{u}rg Fr\"{o}hlich$^3$ and Eugene V. Sukhorukov$^1$}
\affiliation{$^1$D\'epartement de Physique Th\'eorique, Universit\'e de Gen\`eve, CH-1211 Gen\`eve 4, Switzerland}
\affiliation{$^2$Department of Physics, Kyiv National University, 03022 Kyiv, Ukraine}
\affiliation{$^3$Institute of Theoretical Physics, ETH H\"{o}nggerberg, CH-8093 Zurich, Switzerland}
\affiliation{$^4$Bogolyubov Institute for Theoretical Physics, Kiev 03780, Ukraine}

\begin{abstract}
We propose direct experimental tests of the effective models of fractional quantum Hall edge states.
We first recall a classification of effective models based on the requirement of anomaly cancellation
and illustrate the general classification with the example of a quantum Hall fluid at filling factor $\nu = 2/3$.
We show that, in this example, it is {\em impossible} to describe the edge states with only one chiral channel
and that there are several inequivalent models of the edge states with two fields. We focus our attention
on the four simplest models of the edge states of a fluid  with $\nu=2/3$ and evaluate charges and scaling
dimensions of quasi-particles. We study transport through an electronic Mach-Zehnder interferometer and
show that scaling properties of the Fourier components of Aharonov-Bohm oscillations in the current provide
information about the electric charges and  scaling dimensions of quasi-particles. Thus Mach-Zehnder interferometers
can be used to discriminate between different effective models of fluids corresponding to the same filling
factor. They therefore can be used to test fundamental postulates underlying the low-energy effective theory
of edge states. An important ingredient of our analysis is the tunneling Hamiltonian of quasi-particles, the form
of which is discussed in detail.
\end{abstract}


\maketitle

\section{Introduction.}
\label{sec:motivations}

\begin{figure}[h]
\begin{center}
\epsfig{width=5cm,height=3cm,figure=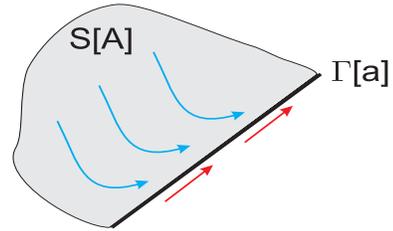}
\caption{(Color online) Illustration of the holography in a QH system.
The Hall current in the bulk (blue arrows) has an anomaly at the edge (i.e., it is not conserved).
This anomaly must be canceled by the anomaly of the edge current (red arrows). \cite{inflow}
Therefore, there is an anomalous boundary action $\Gamma[{\bf a}]$ that is constrained
by the requirement of anomaly cancellation in the bulk effective action $S[{\bf A}]$.}
\label{anomal}
\end{center}
\vskip -3mm
\end{figure}

The quantum Hall effect \cite{QHE,Datta} (QHE) is a fascinating example of macroscopic quantum phenomena. It
continues to attract much attention among experimental and theoretical physicists. Its large-scale physics
is governed by the requirement of anomaly cancellation at the boundary of the system. It provides an example
of the so called ``holographic principle'', \cite{hologr,Witten}
which means that the physics of  the system confined to some region is encoded in the physics of the degrees of
freedom at the boundary of this region (see Fig.\ \ref{anomal}). Understanding the physics of the quantum Hall
(QH) edge states is therefore important for an understanding of the QHE in general.

In the theoretical description of incompressible quantum Hall fluids, the ``holographic principle'' manifests
itself in the presence of {\em chiral} edge channels in the low-energy effective theory. \cite{QHE}
These boundary channels are thought to be described by chiral conformal field theory.\cite{Wen,Frol}
The possible structure of this description is highly constrained by the requirements of locality, the gauge invariance
(charge conservation), and the presence of excitations describing electrons (i.e., with quantum numbers of an electron
or hole) in the spectrum. These requirements allow one to classify possible effective low-energy models for all
observed filling factors. \cite{Fr-abel,Fr-non-abel} However, without taking into account microscopic
properties of a particular incompressible QH state, the requirements mentioned above usually do not determine
the low-energy effective theory uniquely.
Already in examples of incompressible Hall fluids corresponding to simple fractions,
such as $\nu=2/3$, $2/5$, etc., there are several physically inequivalent models satisfying all the requirements
even if one limits one's attention to models with the smallest possible number of edge channels. This situation
calls for experimental tests of the theory.

There are several proposals to test the QH edge physics. Concrete attempts for such tests have been made
in three directions: measurement of the electric charge, of the statistical phase, and of the
{\em scaling dimension} of excitations (quasi-particles and electrons). The scaling dimension of the
electron field operator may be tested via the I-V curve at a tunnel junction. \cite{wen-iv} This idea
has been experimentally implemented in Refs.\ [\onlinecite{Chang-West,Gr-Chang}]. The results of these experiments
caused an extensive discussion of the well known ``2.7 problem''. \cite{podborka} The interpretation
of the experiments described in Refs.\ [\onlinecite{Chang-West,Gr-Chang}] is not straightforward, and one may need
to take into account the existence of compressible strips \cite{strips} at the edge, of disorder, \cite{PKF} and
of electron-phonon interactions. \cite{phonon} The charge of quasi-particles may be probed by measuring the Fano
factor of the tunneling current. \cite{noise-prop} Currently, fractional charges have been observed in several
experiments \cite{fract-charge-meas, quarter-charge} at different filling factors. Several proposals have
been made to use the Fabry-Perot \cite{Goldman} (FP) and the Mach-Zehnder \cite{mz1} (MZ) electronic interferometers,
which utilize QH edge channels, in place of optical beams,
for measurements of fractional charge and of anyon statistics of quasi-particles. \cite{Safi}$^-$\cite{Averin}
%However, these proposals do not discuss the question of how to  distinguish different effective models.

In this work we propose an experiment that would allow one to find out which effective model describes a particular
filling factor. The idea is to use an MZ interferometer in order to measure simultaneously the charge and the scaling
dimension for each species of quasi-particles that may tunnel through a quantum point contact (QPC).
Aharonov-Bohm (AB) oscillations in an MZ interferometer have recently been investigated experimentally
in Refs.\  [\onlinecite{Heiblum2}-\onlinecite{mz7}] and showed surprising behavior. This has been addressed in
several theoretical works. \cite{Sukh-Che,Chalker,Neder,Sim,our}
The main result of the present work, Eqs.\ (\ref{curr-v}) and (\ref{curr-t}) describing
the AB-oscillating contribution to the current through an MZ interferometer at low and high
temperatures, shows that the Fourier spectrum of the current as a function of the flux can be used to extract
the scaling dimensions and charges of quasi-particles.  As an example,
we consider the well observed filling factor $\nu = 2/3$ in some detail. We present possible effective
models for this filling factor and discuss how they can be distinguished from each other with the help
of their spectra of scaling dimensions.

An important property of the $\nu=2/3$ state is that all minimal models of its edge degrees of freedom contain
two edge channels. A similar situation is encountered at $\nu=2$, where it has been shown that the {\em long-range}
Coulomb interaction between the channels leads to some universalities. \cite{our} In this paper we show that
Coulomb interactions fix a freedom in the choice of the edge Hamiltonian, so that scaling dimensions are fully
determined by the matrix (\ref{phase-cond}) of statistical phases of electrons (see Sec.\ \ref{coulomb}).

We start our paper by recalling the effective theory of QH edges and the classification of possible models.
In Sec.\ \ref{bulk-edge} we formulate  general requirements that any model must satisfy.
We illustrate the implementation of these  requirements with the example of fluids with $\nu = 1/(2k+1)$,
where a simple hydrodynamic approach can be used, and discuss limitations of this approach.
In Sec.\ \ref{sec:multi-field-edge}, we discuss general multi-channel edge models and recall the construction
of local excitations and of their correlation functions. In Sec.\  \ref{sec:simplest-two-field} we explicitly
determine the spectrum of scaling dimensions for the most plausible (minimal) models corresponding to $\nu=2/3$.
Section \ref{coulomb} is devoted to an analysis of the role of Coulomb interactions.
Finally, in Sec.\ \ref{mz-sect}, we investigate transport trough an electronic MZ interferometer and show how
scaling dimensions of excitations can be extracted from AB oscillations of the current through the interferometer.
Our form of the tunneling Hamiltonian, which is an important ingredient of the theory, is thoroughly discussed in
Appendix \ref{choice}.

\section{Effective theory of QH edges}
\label{bulk-edge}

The effective theory presented in this section provides a description of the low-energy physics of a QH edge.
While the correct model of edge states may depend on microscopic details of a two dimensional electron gas, there are general
physical requirements that greatly reduce the number of relevant models.\cite{Wen,Frol} These requirements are as follow:
\begin{itemize}
\item {\em Cancellation of anomaly.} It is well known that the Chern-Simons theory of an incompressible quantum
Hall state is anomalous, i.e., in the presence of a boundary, its gauge variation is given by a non-vanishing boundary term.
The effective model of edge states has to be chosen in such a way as to cancel the anomaly of the bulk action
in order for the complete theory to be gauge invariant.
\item {\em Existence of an electron operator.} A two-dimensional electron gas consists of electrons. Thus, on a microscopic
level, the quantum Hall state is described by an electron wave  function. This implies that, in the effective edge theory,
there should exist at least one local operator describing the creation or annihilation of an electron or hole, i.e., with
a charge $e$ and Fermi statistics.
\item {\em Single-valuedness in the electron positions.} Similarly, because the QH state describes electrons, its wave function
must be single-valued in  the electron positions, irrespective of whether quasi-particles are present. As a consequence,
in the effective theory the mutual statistical phase of a quasi-particle and an electron must be an integer multiple of
$\pi$.\cite{footnote2}
\end{itemize}
Below we use these requirements to construct the most simple effective models
of QH edge states and to classify various multi-field models in Sec.\ \ref{sec:multi-field-edge}.

\subsection{Chern-Simons theory and gauge anomaly}
\label{Chern-Simons}

First of all, in an incompressible quantum Hall fluid the electric current density, ${\bf j}$, in the bulk of the system
\cite{Fr-abel} is related to the electromagnetic potential ${\bf A}$ by Hall's law
\begin{equation}
j^\mu = \sigma_H \epsilon^{\mu\nu\lambda} \partial_{\nu} A_{\lambda},
\label{eq:1}
\end{equation}
where the constant $\sigma_H = \nu/2\pi$ is the Hall conductivity and the rational number $\nu$ is the filling factor;
(here and below, we use units where $e=\hbar=1$, and the Einstein summation convention is followed,
unless specified otherwise).  The effective action that leads to Hall's law (\ref{eq:1}) via
$j^\mu= \delta S_{\rm cs}/\delta A_\mu$ is the  three-dimensional Chern-Simons action
\begin{equation}
S_{\rm cs} = \frac{\sigma_H}2 \int_D d^3 r\, \epsilon^{\mu\nu\lambda}
A_\mu \partial_\nu A_\lambda,
\label{eq:3}
\end{equation}
where $D$ is the product of the time axis and some spatial domain.

Action (\ref{eq:3}) is \emph{anomalous}, i.e., it has a nonvanishing gauge variation in the presence of a boundary.
Indeed, gauge transforming the potential $A_\lambda\to A_\lambda+\partial_\lambda f$ in Eq.\ (\ref{eq:3}) by an
arbitrary gauge function $f(r)$ and integrating by parts, we obtain the variation of the action:
\begin{equation}
\delta S_{\rm cs} = \frac{\sigma_H}2 \int_{\partial D} d^2 r\,\epsilon^{\nu\lambda}
\partial_\nu A_\lambda f.
\label{eq:3-anomaly}
\end{equation}
This anomaly originates from the fact that the current (\ref{eq:1}) is not conserved at the  boundary:
$\partial_\mu j^\mu\neq 0$, for $r\in\partial D$. Indeed, taking the derivative of Eq.\ (\ref{eq:1}), we find that
\begin{equation}
\partial_\mu j^\mu = \sigma_H\epsilon^{\mu\nu\lambda}\partial_\nu A_\lambda\partial_\mu\theta_D,
\label{anomaly1}
\end{equation}
where the function $\theta_D$ takes values $\theta_D = 1$ and $\theta_D = 0$ inside and outside the domain
$D$, respectively.

The anomaly must be compensated by boundary degrees of freedom coupled to the electromagnetic field.
Namely, the total effective action, after the boundary fields are integrated out, is given by a sum of two terms,
\begin{equation}
S_{\rm tot}[{\bf A}] = S_{\rm cs}[{\bf A}]+\Gamma[{\bf a}],
\label{tot}
\end{equation}
where ${\bf a}$ is the electromagnetic field at the boundary,
${\bf a} = {\bf A}|_{{\partial D}}$, and $\Gamma[{\bf a}]$ is an anomalous action at the edge. \cite{Jackiw}
The anomaly in $\Gamma$ must be such that, under a gauge transformation $a_\lambda\to  a_\lambda+\partial_\lambda f$,
this action acquires a variation that cancels exactly the one of the bulk action, see (\ref{eq:3-anomaly}):
$\delta \Gamma=-\delta S_{\rm cs}$. Under this condition, $S_{\rm tot}[{\bf A}]$ is gauge-invariant.  Consequently,
the edge current ${\bf J}$, defined as $J^\mu =\delta(S_{\rm cs}+\Gamma)/\delta a_\mu$, is anomalous, with divergence
given by
\begin{equation}
\partial_\mu J^\mu = \sigma_H\epsilon^{\mu\nu}\partial_\mu a_\nu
\label{anomaly2}
\end{equation}
This divergence cancels the divergence (\ref{anomaly1}) of the bulk current.
Below we discuss various models that incorporate these general ideas,
starting from a simple hydrodynamical model.

\subsection{Hydrodynamics of incompressible edge deformations}
\label{hydrodynamics}

\begin{figure}[thb]
\epsfxsize=7cm
\epsfbox{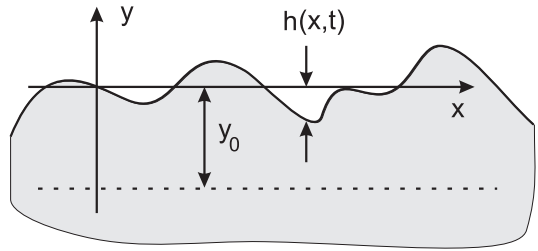}
\caption{Deformations of the boundary (thick line) of
an incompressible QH fluid (shown in gray) are parameterized by the
function $y=h(x,t)$. The auxiliary boundary, where bulk and edge currents match,
is indicated by the dashed line $y=-y_0$.}  \label{deformations}
\vskip -2mm
\end{figure}

Next, we recall some arguments in Ref.\ [\onlinecite{wen-text}], with some modifications taking into account the physics related
to long-range Coulomb interactions. Edge excitations can be viewed as deformations of the boundary of an incompressible QH liquid
caused by the bulk current flowing towards the edge (see Fig.\ \ref{deformations}). We parametrize these deformations by a
function $y=h(x,t)$ and consider a low-energy limit, so that $h\ll\ell_h$, where $\ell_h$ is the characteristic wave length of
the deformations. Introducing an auxiliary boundary at a distance $y_0$ from the edge, with $\ell_h\gg y_0\gg h$,
we represent the edge current $J\equiv J_x$ as the integral $J=\int_{-y_0}^{h}dy j_{x}$, and, for the accumulated charge density
at the edge, $\rho\equiv J_t$, we write $\rho=n_0h$, where $n_0=\nu B/2\pi c$ is the density of the QH liquid and $B$ is
the magnetic field value.

An unpleasant aspect of this approach is that the edge current explicitly depends on the auxiliary cutoff at $y=-y_0$.
However, we show that the resulting equation of motion for the edge deformations, $h(x,t)$, does not contain this cutoff,
and hence the edge current can be redefined to depend on $h$ only. Indeed, charge conservation implies that
$\partial_t\rho+\partial_xJ=j_y$, where the bulk current density $j_y$ is taken at the boundary $y=-y_0$.
Using Eq.\ (\ref{eq:1}) and fixing the gauge $a_x =0$, we write $J=\sigma_H[\varphi(x,h)-\varphi(x,-y_0)]$, where $\varphi(x,y)$
is the total electrostatic potential in the plane of the QH fluid. Substituting this expression for the current in the continuity
equation, we observe that the terms containing $y_0$ cancel, and the equation of motion takes the form
$\partial_t\rho+\sigma_H\partial_x\varphi=0$. Finally, we split the potential $\varphi$ into two parts and write
$\partial_x\varphi=\partial_x\varphi_h- \tilde E_x$, where $\varphi_h$ is the potential at the edge caused by its deformation,
and $\tilde E_x$ is the external electric field evaluated at $y=h$. While $\tilde E_x$ depends on $h$ in general,
in the low-energy limit it can be taken at $y=0$, to leading order in $h$. The equation of motion then reads,
\begin{equation}
\partial_t\rho+\sigma_H\partial_x\varphi_h=\sigma_H\tilde E_x.
\label{motion}
\end{equation}
After a redefinition of the edge current, $J=\sigma_H\varphi_h$, the cutoff parameter $y_0$ is gone.

In order to close the equation of motion, we need a relation between the deformation $h$ and the potential $\varphi_h$.
In the long wave-length limit, we can simply write $\varphi_h=-\partial_\rho H=-(1/n_0)\partial_h H$, where $H$ is the density
of electrostatic energy at the edge. This leads to an equation that is, in general, non-linear in the field $\rho$.
Passing to a low-energy limit, this equation can be linearized, and we arrive at the result:
\begin{equation}
\partial_t\rho-v\partial_x\rho=\sigma_H\tilde E_x,
\label{linear-eq}
\end{equation}
where $v=(\sigma_H/n_0^2)\partial^2_hH(0)$ is the group velocity of the edge excitations.
For a stable QH liquid, $\partial^2_hH(0)$ is positive, and this equation describes the
propagation of chiral excitations.

There are two contributions to the electrostatic energy: one is due to the confining potential at the edge and
the second one is due to Coulomb interactions. Consequently, the group velocity of edge  excitations can be
written as a sum of two terms,
\begin{equation}
v=cE/B+\sigma_HV,
\label{velocity}
\end{equation}
where the first term is the velocity of drift in the electric field
$E$ at the edge of the QH liquid, and the second term is proportional
to the integral $V=\int dx'U_C(x-x')$ of the Coulomb interaction potential $U_C$ at the edge.
This integral is logarithmically divergent and has to be cut off at the distance, $d$, to the metallic gate
and at the microscopic width of the edge $a$, so that $V\sim\ln(d/a)$.

Here an important remark is in order. The first term in Eq.\  (\ref{velocity}) may be interpreted as a bare
velocity, $v_0$, of excitations. Restoring physical units, it can be estimated as $v_0\sim  (eE/\hbar)l_{B}^2$,
where $l_B$ is the magnetic length. The ratio $\alpha :=\sigma_H/v_0\sim e^2/\hbar v_0$ plays the role of
a dimensionless interaction constant. Depending on the confinement at the edge, it is always larger than 1,
and, in a typical  experiment,\cite{mz1,Heiblum2,Basel,Glattli1,mz7} $\alpha\geq 10$, which may justify
the hydrodynamical model considered here.
%However, we have neglected the correlation contribution to the total
%energy $H$, which, in contrast to the contribution from the confining potential, is not small. This might be
%one of the reasons for the failure of the hydrodynamical model when applied to QH fluids at certain filling factors;
%see the discussion below.
Moreover, the long-range character of the Coulomb interaction and the fact that $d\gg a$ leads to
a large parameter $V$. As a result, the hydrodynamical charged mode is always present in the spectrum and determines
the scaling dimension in non-chiral models, as we demonstrate in Sec.\ \ref{coulomb}.

\subsection{Quantization of edge excitations}
\label{quantization}

In order to quantize edge excitations, we consider the total electrostatic energy density $H(\rho)=(v/2\sigma_H)\rho^2$
as a Hamiltonian that generates the homogeneous version of the equation of motion (\ref{linear-eq}),
see Ref.\ [\onlinecite{wen-text}].
This equation is diagonal in Fourier space, $\partial_t\rho_k-ivk\rho_k=0$. We therefore write the Hamiltonian as
\begin{equation}
H_k=\frac{v}{\sigma_H}\rho_k\rho_{-k},
\end{equation}
where $k>0$. We identify the ``momentum'' with $P_k=\rho_k$ and the ``coordinate'' with
$X_k=i\rho_{-k}/\sigma_Hk$, so that the equations of motion take the form $\partial_t X_k=\partial H_k/\partial P_k$
and  $\partial_t P_k=-\partial H_k/\partial X_k$. Then the canonical
commutator $[X_k,P_{k'}]=i\delta_{kk'}$ leads to the commutator $[\rho(x),\rho(x')]=
i\sigma_H\partial_x\delta(x-x')$ in real space.

Next, we construct an electron operator. For this purpose it is convenient to represent
the charge density in terms of a field $\phi(x)$,
\begin{equation}
\rho(x)=\frac{\sqrt{\nu}}{2\pi}\;\partial_x\phi(x),
\label{charge-density}
\end{equation}
with commutation relations
\begin{equation}
[\phi(x'),\phi(x)]=i\pi\,{\rm sgn}(x-x').
\label{normalization}
\end{equation}
Here, and in the following, we use the term ``filling fraction'', $\nu$, and ``Hall conductivity'',
$\sigma_H$, synonymously; but we always mean the latter.
Then the electron operator takes the form
\begin{equation}
\psi = e^{iq\phi}
\label{el-1}
\end{equation}
of a local vertex operator; see, e.g., Ref.\ [\onlinecite{boson-textbook}].
For this operator to describe the creation and annihilation of an electron or hole, we require that
\begin{equation}
[Q_{\rm em},e^{iq\phi}] = e^{iq\phi},
\label{charg-1}
\end{equation}
where $Q_{\rm em} = \int dx \rho = (\sqrt{\nu}/2\pi)\int dx \partial_x\phi$ is the total
electric charge at the edge. This requirement implies that the charge of an electron
is equal to $-1$. Using the commutation relations (\ref{normalization}) we find that
\begin{equation}
q =1/\sqrt{\nu}.
\label{result-for-q1}
\end{equation}
In addition, an electron operator (\ref{el-1}) must obey fermionic commutation relations.
Applying the Baker-Campbell-Hausdorff formula, we find that
$e^{iq\phi(x)}e^{iq\phi(x')} = e^{i\pi  q^2}e^{iq\phi(x')}e^{iq\phi(x)}$.
Using Eq.\ (\ref{result-for-q1}) and imposing Fermi statistics, we conclude that
\begin{equation}
e^{i\pi/\nu} = -1.
\end{equation}
This implies that the filling factor is given by  $\nu = 1/m$, where $m$ is an odd integer number.
In the Sec.\  \ref{sec:multi-field-edge}, we show that this limitation can be overcome by constructing
a multi-channel edge model.

According to the third principle formulated at the beginning of this section, the theory may describe
quasi-particles with the vertex operators $e^{ip\phi}$ that must be local relative to the electron operator
$e^{iq\phi}$. Thus the statistical phase, $\theta$, of such quasi-particles with respect to an electron  has
to be an integer multiple of $\pi$. Using again the commutation relation (\ref{normalization}),
we arrive at the result that
$$
\theta = \pi p\cdot q = \pi n,
$$
and the quasi-particle operator takes the form:
\begin{equation}
\psi_n = e^{in\sqrt{\nu}\phi(x)},
\label{psi-n}
\end{equation}
where $n$ is an integer. Such operators describe Laughlin quasi-particles. \cite{Laugh}
The correlation functions of quasi-particle operators may be calculated easily, with the result
$\langle 0|\psi_n^\dag(x,t)\psi_n(0,0)|0\rangle = (x+vt)^{-\nu n^2}$, where $|0\rangle$ denotes
the ground state of a quantum Hall fluid with filling fraction $\nu$.
Taking into account that $\nu = 1/m$, the properties of the operators (\ref{psi-n}) are as follows:
they carry a charge $q(n) = n/m$ and have the scaling dimensions $\Delta(n) = n^2/m$. Thus, for an
elementary quasi-particle with charge $1/m$, we have that $\Delta_{\rm min}=1/m$, and, for an electron,
$\Delta_{\rm el} = m$.

\subsection{Gauge-invariant formulation}
\label{gauge-invariant}

%We remind that the above derivations have been carried out in a fixed gauge $a_x=0$.
In this section we reformulate the theory of edge excitations presented above
in a gauge-invariant form suitable for a generalization to multi-channel fluids
considered in Sec.\ \ref{sec:multi-field-edge}. We first rewrite the action
$S=\int dt \sum_{k>0}[P_k\partial_t X_k-H_k]$ in the linear approximation as
\begin{equation}
S[\phi]=\frac{1}{4\pi}\int dtdx [\partial_t\phi\partial_x\phi -  v(\partial_x\phi)^2
+2\sqrt{\nu}\phi\tilde E_x],
\label{linear-action}
\end{equation}
where we have included a term describing the coupling to an electric field $\tilde E_x$.
This action can  easily be generalized to nonlinear edge modes by  replacing the term
$(v/4\pi)(\partial_x\phi)^2$ with the full
Hamiltonian $H(\rho)$.
%However, the dynamics in general is not  integrable.

Next, we replace derivatives $\partial_\mu\phi$ in action
(\ref{linear-action}) with their gauge-invariant form
\begin{equation}
D_\mu\phi\equiv\partial_\mu\phi+\sqrt{\nu}a_\mu,
\label{derivatives}
\end{equation}
and integrate the last term by parts using the relation $\tilde  E_x=\epsilon^{\mu\lambda}\partial_\mu a_{\lambda}$.
We then arrive at the following action
\begin{multline}
S[\phi]=\frac{1}{4\pi}\int dtdx [D_t\phi D_x\phi - v(D_x\phi)^2] \\+ \frac{\sqrt{\nu}}{4\pi}\int dtdx
%\phi\tilde E_x
\epsilon^{\mu\lambda} a_\mu \partial_\lambda\phi.
\label{invariant-action}
\end{multline}
It is easy to check that by fixing the gauge $a_x=0$, one returns to action (\ref{linear-action}).

The first term in action (\ref{invariant-action}) is invariant under the gauge transformation
$a_\mu\to a_\mu+\partial_\mu f$, $\phi\to \phi-\sqrt{\nu}f$. The second term yields the gauge variation
$\delta S[\phi] = -(\nu/4\pi) \int dtdx \epsilon^{\mu\lambda} \partial_\mu a_\lambda f$,
i.e., the edge action has the desired anomaly: It exactly cancels the anomaly (\ref{eq:3-anomaly}) of
the bulk action. Thus, the effective theory described by the total action $S_{\rm tot}=S_{\rm cs}[{\bf A}]+S[\phi]$
is gauge-invariant. The boundary effective action $\Gamma[{\bf a}]$ in Eq.\ (\ref{tot}) is obtained by integrating out
the field $\phi$.

One may then check that the gauge-invariant edge current has the correct anomalous divergence (\ref{anomaly2}).
To see this, we take a variational derivative of the total action with respect to the boundary potential ${\bf a}$.
This yields the following expression for the edge current:
\begin{equation}
J_t=\frac{\sqrt{\nu}}{2\pi}D_x\phi,\quad J_x=-\frac{\sqrt{\nu}}{2\pi}v D_x\phi.
\label{edge-current}
\end{equation}
In a gauge where $a_x=0$, this expression reproduces definition (\ref{charge-density}) of the charge density
as well as the definition of electron operator (\ref{el-1}). Indeed,
expression (\ref{edge-current}) for the edge current follows from point-splitting of the operator
$\psi^\dag\psi$ in the presence of an electromagnetic field.
Note that the current satisfies the relation $J_x=-vJ_t$, which exhibits its chiral nature.
Finally, by varying action (\ref{invariant-action}) with respect to $\phi$, we obtain the equation of motion
for $\phi$, which is used to evaluate the divergence of the current (\ref{edge-current}). We then arrive at
Eq.\ (\ref{anomaly2}), i.e., the edge current has the desired anomalous divergence.

We conclude that the hydrodynamical model, when applied to QH states with $\nu=1/m$ where
$m$ is an odd integer,  satisfies all the requirements formulated at the beginning of this section.
In Sec.\ \ref{sec:multi-field-edge}, we show that by considering more than one bosonic mode at the edge of a QH liquid,
one can construct effective edge models for general filling factors.

\section{Multi-channel edge models}
\label{sec:multi-field-edge}

As shown in Sec.\ \ref{quantization}, a single-channel hydrodynamical model of the QH edge cannot describe
all observed  filling fractions. We therefore consider more general multi-channel edge models. A natural
generalization of single-field action (\ref{invariant-action}) to many fields is given by
\begin{multline}
S[\phi_i] = \frac{1}{4\pi}\sum_i \int dt dx [\sigma_iD_t\phi_iD_x\phi_i - v_i(D_x\phi_i)^2] \\ + \frac{1}{4\pi}\sum_i\int dt dx
[Q_i\epsilon^{\mu\lambda}a_\mu\partial_\lambda\phi_i],
\label{act-2}
\end{multline}
where $\sigma_i = \pm 1$ encodes the chirality of the $i^{\rm th}$ channel, $v_i$ is the propagation speed, and $Q_i$
is the constant of electromagnetic coupling of the field $\phi_i$. The covariant derivatives are defined by
$D_\mu\phi_i = \partial_\mu\phi_i+\sigma_iQ_ia_\mu$.
We emphasize that any quadratic gauge-invariant action for chiral bosons can be brought to
the unique form (\ref{act-2}) by redefining the fields.
Here we consider the general case with different
propagation speeds, $v_i$, for  different edge modes, because recent experiments \cite{Heiblum2,Basel,Glattli1,mz7}
show that this can occur.

The requirement of anomaly cancellation for edge action (\ref{act-2}) implies that
\begin{equation}
\sum\limits_i \sigma_iQ_i^2 = \nu.
\label{canc-2}
\end{equation}
One can see that, in contrast to a single-channel edge where the last term in action (\ref{invariant-action})
is uniquely fixed by the Hall conductivity, in the multi-channel situation only the ``length'' of the vector $Q_i$
is fixed to be $\sqrt{\nu}$, while, at this point, its direction is still arbitrary.

\subsection{Kinematics of edge models}
\label{classification-edge}

In order to check the second physical requirement, the existence of excitations with the quantum numbers of electron,
we consider a general vertex operator
\begin{equation}
\psi = \exp\Big(i\sum_j\!q_j\phi_j\Big),
\label{op-2}
\end{equation}
where $q_j$ are some constants.
Taking into account commutation relations
\begin{equation}
[\partial_x\phi_i(x,t),\phi_j(x',t)] = -2\pi i\sigma_i\delta_{ij}\delta(x-x'),
\label{comut-2}
\end{equation}
which follow from Eq.\ (\ref{act-2}),
we find that the statistical phase of operator (\ref{op-2}) is given by:
\begin{equation}
\theta = \pi\sum_i \sigma_iq_iq_i.
\label{ph-def}
\end{equation}
The electric charge operator is given by $Q_{\rm em} = (1/2\pi)\sum_iQ_i\int dx\partial_x\phi_i $,
in accordance with Eq.\ (\ref{act-2}). Therefore, one finds with the help of Eq.\ (\ref{comut-2}) that
the charge of field operator (\ref{op-2}) is given by
\begin{equation}
Q_{\rm em} = \sum_i\sigma_iQ_iq_i.
\label{ch-def}
\end{equation}

Here, as in the integer QHE, it is possible to have several electron operators differing from each
other by some quantum numbers. The origin of these quantum numbers is discussed in Appendix \ref{cs-bulk}.
We label different electron operators by an additional index $\alpha$,
\begin{equation}
\psi_\alpha = \exp\Big(i\sum_j\!q_{\alpha j}\phi_j\Big),
\label{exc-el}
\end{equation}
and assume that the number of electrons coincides with the number of channels.\cite{footnote-ckm}
All electron fields must have a unit charge, which implies that
\begin{equation}
\sum_i\sigma_iQ_iq_{\alpha i}= 1,
\label{charg-cond}
\end{equation}
and appropriate relative statistical phases, $\pi K_{\alpha\beta}$, compatible with
relative locality and Fermi statistics. This implies that the numbers
\begin{equation}
K_{\alpha\beta} = \sum_i \sigma_iq_{\alpha i}q_{\beta i}
\label{phase-cond}
\end{equation}
must be integers and that, for $\alpha = \beta$, these numbers must be odd integers.

\begin{figure}[h]
\begin{center}
\vskip -3mm
 \epsfig{width=7cm,height=6cm,figure=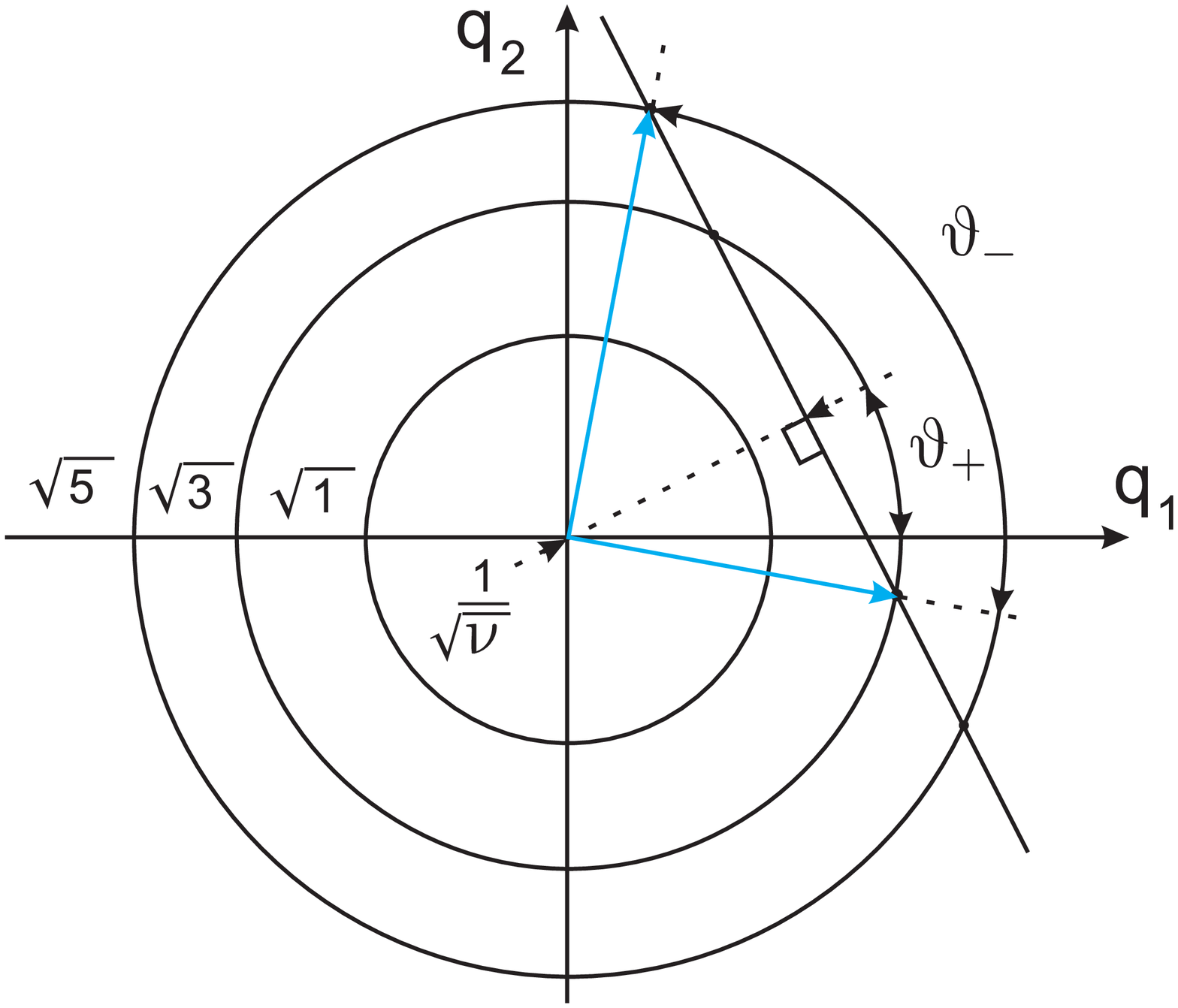}
\caption{(Color online) Schematic illustration of the conditions for electron operators in a chiral two-field model.
The requirement for the statistical phase of an electron operator to be fermionic is
$q_{\alpha 1}^2 +q^2_{\alpha 2} = 2k+1$. This means that the end points of vectors
$\mathbf{q}_\alpha \equiv \{q_{\alpha i}\}$ (drawn in blue) lie on the circle of radius $\sqrt{2k+1}$.
The condition of unit charge, $Q_1q_{\alpha 1}+Q_2q_{\alpha 2}= 1$, implies that the end points
of vectors $\mathbf{q}_\alpha$ lie on the line perpendicular to the vector $\mathbf{Q}\equiv\{Q_i\}$.
The length of this vector is fixed by the anomaly cancellation condition (\ref{canc-2}), namely
$|\mathbf{Q}| = \sqrt{\nu}$. Therefore the distance from the line through the end points of the vectors
$\mathbf{q}_\alpha$ to the origin is fixed to be $1/\sqrt{\nu}$. We denote the angle between
$\mathbf{Q}$ and the $q_1$ axis by $\vartheta_+$, and the angle between the electron vectors by
$\vartheta_-$.} \label{latt-fig}
\end{center}
\vskip -6mm
\end{figure}

In Fig.\ \ref{latt-fig} we schematically illustrate conditions (\ref{charg-cond})
and (\ref{phase-cond}) for the simple example of two channels with the same chiralities.
One sees that in contrast to the single-channel case, in multi-channel models there is a
freedom in choosing electron operators, even if the coupling constants $Q_i$ are fixed.
This freedom implies that different microscopic QH wave-functions may
lead to the same action in the low-energy limit. In this case the low-energy projections
of electronic operators may in principle be different.
In fact, multi-channel models are fully determined by the numbers $q_\alpha^i$, while the values of
coupling constants $Q_i$ can be obtained by solving Eq.\ (\ref{charg-cond}):
\begin{equation}
Q_i = \sigma_i\sum_\alpha q^{-1}_{i\alpha}.
\label{q-q}
\end{equation}
The physical requirements for the effective theory can therefore be formulated as constraints on the
$q$-matrix. Namely, the requirement that elements of the matrix $K$ given by Eq.\ (\ref{phase-cond})
are integer numbers has to be accompanied by the condition that
\begin{equation}
\sum_{\alpha,\beta}K^{-1}_{\alpha\beta} = \nu,
\label{cond-main}
\end{equation}
which follows from the requirement of anomaly cancellation (\ref{canc-2}) and from equation (\ref{q-q}).

Note that we have reformulated the constraints on the matrix $q$ as constraints on the matrix $K$.
In Sec.\ \ref{sec:excitations-theory} we show that the {\em kinematic} information about an effective model is encoded
in the matrix $K$. More precisely, the spectra of statistical phases and charges of quasi-particles
are entirely determined by $K$. For every filling factor this matrix takes values from a discrete set.
For instance, in two-channel models this corresponds to the discrete set of choices of lengths of
electronic vectors $\mathbf{q}_\alpha$ and their relative angle $\vartheta_-$ (see Fig.\ \ref{latt-fig}).
From the relation (\ref{phase-cond}) it follows that the remaining freedom in the matrix $q$ for a
given matrix $K$ is the angle $\vartheta_+$ of the simultaneous rotation of two vectors
$\mathbf{q}_\alpha$. In Sec.\ \ref{sec:excitations-theory} we show that the {\em dynamical} properties of the model,
such as the correlation functions, are not determined by the matrix $K$ only, but depend on the
whole matrix $q$, e.g., on the angle $\vartheta_+$ in the case of two fields.

\subsection{Local excitations}
\label{sec:excitations-theory}

Next, we determine all quasi-particle operators in accordance with the requirement that they have
integer statistical phases relative
to all electron fields (\ref{exc-el}). Quasi-particle operators are vertex operators of the form
\begin{equation}
\psi = \exp\Big(i\sum_j\!p_j\phi_j\Big) .
\label{exc}
\end{equation}
Their statistical phases relative to electronic fields are given by
\begin{equation}
\theta_{pq_\alpha} = \pi \sum_i\sigma_ip_i q_{\alpha i}  = \pi n_\alpha.
\label{ekv}
\end{equation}
The numbers $n_\alpha$ must be integers. The solution of Eq.\ (\ref{ekv}),
\begin{equation}
p_i = \sigma_i\sum_\beta q^{-1}_{i\beta }n_\beta,
\label{alow}
\end{equation}
is a linear combination with integer coefficients. Therefore, the whole set of allowed quasi-particle operators
forms a lattice, which is dual to the lattice spanned by electronic vectors $q_{\alpha i}$
(see Appendix \ref{sec:frohlich-theory} and Ref.\ [\onlinecite{Fr-abel}] for a detailed discussion of this point).

It is interesting to note that the statistical phase and the charge of a quasi-particle operator labeled by the
numbers $n_\alpha$ can be expressed solely in terms of the matrix $K$. For the statistical phase we have that
\begin{equation}
\frac{\theta}{\pi} = \sum_i\sigma_i p_ip_i = \sum_{\alpha,\beta}n_\alpha K^{-1}_{\alpha\beta}n_\beta.
\label{phas-v}
\end{equation}
It also follows from Eqs.~(\ref{q-q}) and (\ref{alow}) that the charge of the operator in (\ref{exc}) is given by
\begin{equation}
Q_{\rm em} = \sum_i\sigma_iQ_ip_i = \sum_{\alpha\beta} K^{-1}_{\alpha\beta}n_\beta.
\label{charg-v}
\end{equation}
Note that the summation over the index $\alpha$ in this equation may be viewed as the multiplication by
the vector $(1,1,\ldots,1)$.

It may happen that different matrices $K$ generate the same set of quasi-particles. This is the case
when corresponding electronic vectors $q_{\alpha i}$ form different bases of the same lattice (see the discussion
in Appendix \ref{sec:frohlich-theory}). An example of such an equivalence is depicted in Fig.\ \ref{latt-eq}.
In the language of matrices $q$, an equivalence is the consequence of the fact that an integral transformation
${q'}_{\alpha i} = \sum_\beta T_{\alpha\beta} q_{\beta i}$ (i.e., one with the elements $T_{\alpha\beta}$
and $T^{-1}_{\alpha\beta}$ being integer numbers) is nothing but an automorphism of the integral lattice
generated by a change of basis. Using definition (\ref{phase-cond}), this equivalence may also be written as
\begin{equation}
K \leftrightarrow K' = TKT^{T},
\label{equiv}
\end{equation}
Since the matrix $T$ transforms an electronic basis, it preserves the charge of an electron.
Taking into account Eq.\ (\ref{charg-v}), this important condition implies that the
matrix $T$ should preserve the vector $(1,1,\dots,1)$.

In conclusion, we propose the following strategy to find inequivalent models for a given filling factor.
First of all, one must find all solutions, $K$, of Eq.\ (\ref{cond-main}) for a given $\nu$, up to
equivalence defined by proper integral transformations (\ref{equiv}). This procedure fixes the kinematic
content of the theory.\cite{footnote3} Second, one must fix those parameters that are not constrained
by the general conditions formulated at the beginning of Sec.\ \ref{bulk-edge}. These parameters are
the propagation speeds, $v_i$, of chiral edge modes. Finally, one should choose an explicit basis,
$\mathbf{q}_\alpha$, of vectors labeling electron field operators and consistent with the chosen matrix $K$.

\subsection{Scaling dimensions of local excitations}
\label{scaling-dimensions}

We conclude this section by presenting the correlation functions of the quasi-particle operators (\ref{exc}).
A detailed calculation of this function is contained in Appendix \ref{corr-f-calc} and yields
\begin{equation}
\langle
0|\psi^\dag(x,t)\psi(0,0)|0\rangle \propto e^{i\varphi_0(\mathbf{n})}\prod_i(x+\sigma_iv_it)^{-\delta_i(\mathbf{n})},
\label{corr-m}
\end{equation}
where the exponents are given by
\begin{equation}
\delta_i(\mathbf{n}) = p_i^2 = \Big[\sum_\alpha q^{-1}_{\alpha i}n_\alpha\Big]^2.
\label{eq:23}
\end{equation}
Here $\mathbf{n}\equiv\{n_\alpha\}$, and $\varphi_0$ is a phase, the exact value of which is discussed in Sec.\ \ref{mz-sect}.

The scaling dimension of the correlation function, defined via its long-time behavior, is given by
\begin{equation}
\Delta(\mathbf{n}) = \sum_i \delta_i(\mathbf{n}).
\label{eq:30}
\end{equation}
Expressed in terms of the $q$-matrix, it reads
\begin{equation}
\Delta(\mathbf{n}) = \sum_i p_i^2 = \sum_{\alpha,\beta} n_\alpha (q q^T)^{-1}_{\alpha\beta} n_\beta.
\label{scaling-d}
\end{equation}
An explicit calculation of $\Delta$ in the non-chiral case with two fields is given in Appendix \ref{correlator}.
The scaling dimensions $\Delta$ are not fully determined by the matrix $K$, while according to Eq.\ (\ref{phas-v}),
the statistical phases
\begin{equation}
\frac{\theta}{\pi} = \sum_i\delta_i\sigma_i
\label{stat-n}
\end{equation}
are given by the matrix $K$. Comparing Eqs.~(\ref{eq:30}) and (\ref{stat-n}) we conclude that $\Delta \geq \theta / \pi$,
where equality holds in a purely chiral case.

\section{Minimal models for $\nu=2/m$, and scaling dimensions of their quasi-particle fields}
\label{sec:simplest-two-field}

In this section we apply the ideas discussed above to the particular case of filling factors $\nu=2/m$.
In Ref.\ [\onlinecite{PKF}] it has been shown that
models with a large number of edge channels may be unstable under the influence of disorder.
To avoid such complications, we limit our analysis to models with the smallest possible number of fields
(see also the discussion at the end of Appendix \ref{cs-bulk}). Moreover, we consider
models with minimal statistical phases of electron field operators, because
they are most relevant physically. \cite{footnote7}

A direct solution of Eq.\ (\ref{cond-main}) is complicated. Fortunately, in Ref.\ [\onlinecite{Fr-abel}],
some general results have been proven for the case where the statistical phases of electron operators are smaller than $7\pi$:
all two-field models for $\nu = 2/m$ are described by matrices $K$ of the following form:
\begin{equation}
K_a = \left(
  \begin{array}{cc}
    a & b \\
    b & a \\
  \end{array}
\right).
\label{k-gener11}
\end{equation}
Equation (\ref{cond-main}) then imposes the following constraint on matrix (\ref{k-gener11}):
\begin{equation}
\nu = \frac{2a-2b}{a^2-b^2} = \frac{2}{a+b}.
\label{ab}
\end{equation}
Thus, for $\nu = 2/m$, the parameters $a$ and $b$ are related by $a+b =m$, where the odd integer $a$ enumerates the models.

For a purely chiral model, the scaling dimensions of correlation functions~(\ref{scaling-d}) are given by the statistical
phases. Therefore, for a $K$-matrix of the form (\ref{k-gener11}), they are given by the expression
\begin{equation}
\Delta(\mathbf{n}) = \frac{1}{a^2-b^2}[a(n_1^2+n_2^2)-2bn_1n_2],
\end{equation}
and the charge of excitations can be evaluated as
\begin{equation}
Q_{\rm em} = \frac{n_1+n_2}{m}.
\end{equation}
For non-chiral models, the expression for the charges of quasi-particles remains the same, while the scaling
dimensions (\ref{scaling-d}) depend on an additional parameter, $\vartheta_+$, (see Appendix \ref{correlator}).
In Sec.\ \ref{coulomb}, we show that, in the limit of strong Coulomb interactions, this parameter takes the universal
value $\vartheta_+ =0$. The scaling dimensions are then given by
\begin{equation}
\Delta(\mathbf{n}) = \frac{1}{b^2-a^2}[b(n_1^2+n_2^2)-2an_1n_2].
\label{non-ch}
\end{equation}

\begin{figure}[ht]
\begin{center}
\vskip -3mm
\epsfig{width=8cm,height=6cm,figure=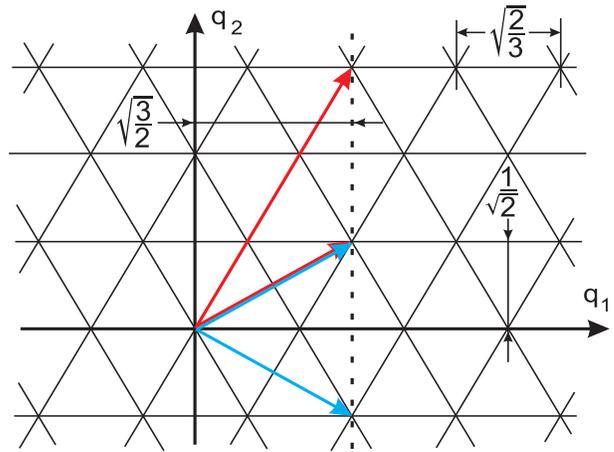}
\caption{(Color online) Illustration of equivalence of two QH lattices. For two channels of different chiralities, the statistical phase
satisfies $\theta/\pi = q_1^2-q_2^2$. Using this fact, one can easily see that the red vectors correspond to $K$-matrix
(\ref{k-videl}) and the blue vectors correspond to the $K$-matrix (\ref{eq:24}). Note that these pairs of vectors are just
different bases of the same lattice (which is dual to the one shown in the figure). Therefore the spectra of statistical
phases in models (\ref{k-videl}) and (\ref{eq:24}) are identical. Finally, the dashed line constrains the charge of the
electrons to be $1$. The fact that all electron vectors lie on the same line implies that the constants $Q_i$ are the same
for both models. This means that the charges $Q_{\rm em} = \sum q_iQ_i$ of excitations in one model coincide with those in
the second model. Here we choose the angle $\vartheta_+ = 0$, in accordance with the conclusion of Sec.\ \ref{coulomb}.}
\label{latt-eq}
\end{center}
\vskip -3mm
\end{figure}

We have already mentioned that any two-field solution of Eq.\ (\ref{cond-main}) corresponds to one of the matrices (\ref{k-gener11}),
up to equivalence described by the transformations (\ref{equiv}). An important example is the $K$-matrix proposed, e.g.,
in Ref.\ [\onlinecite{wen23}]:
\begin{equation}
K = \left(
      \begin{array}{cc}
        1 & 0 \\
        0 & -3 \\
      \end{array}
    \right).
\label{k-videl}
\end{equation}
This matrix describes a non-chiral model of the $\nu=2/3$ state obtained by particle-hole conjugation of the $\nu=1/3$ state.
In this state, the density at the edge first increases to $\nu=1$ and then drops to zero, which implies the presence of two
edge channels with opposite chiralities. Another $K$-matrix for $\nu = 2/3$ state appears in the context of the composite
fermion approach: \cite{jain}
\begin{equation}
K = \left(
  \begin{array}{cc}
    1 & 2 \\
    2 & 1 \\
  \end{array}
\right).
\label{eq:24}
\end{equation}
It turns out that model (\ref{k-videl}) is equivalent to (\ref{eq:24}), in the sense of (\ref{equiv}).
Indeed, one can apply an integral change of variables transforming one $K$-matrix into the other one:
$$
\left(
  \begin{array}{cc}
    1 & 0 \\
    2 & -1 \\
  \end{array}
\right)\left(
         \begin{array}{cc}
           1 & 0 \\
           0 & -3 \\
         \end{array}
       \right)
       \left(
  \begin{array}{cc}
    1 & 2 \\
    0 & -1 \\
  \end{array}
\right) = \left(
  \begin{array}{cc}
    1 & 2 \\
    2 & 1 \\
  \end{array}
\right).
$$
This transformation is of the type of (\ref{equiv}), because it has the property that
$$
\left(
  \begin{array}{cc}
    1 & 0 \\
    2 & -1 \\
  \end{array}
\right)^{-1} = \left(
  \begin{array}{cc}
    1 & 0 \\
    2 & -1 \\
  \end{array}
\right),
$$
and it leaves the vector $(1,1)$ invariant. The equivalence of these two models is illustrated in Fig.\ \ref{latt-eq}.

For $\nu = 2/3$, the matrices (\ref{k-gener11}) with the smallest diagonal elements (i.e., with smallest statistical
phases of electron operators) are the following ones:
\begin{equation}
  K_3 = \left(
  \begin{array}{cc}
    3 & 0 \\
    0 & 3 \\
  \end{array}
\right),\hspace{10pt}
K_5 = \left(
  \begin{array}{cc}
    5 & -2 \\
    -2 & 5 \\
  \end{array}
\right)
\label{eq:25}
\end{equation}
and
\begin{equation}
K_1 = \left(
  \begin{array}{cc}
    1 & 2 \\
    2 & 1 \\
  \end{array}
\right),\hspace{10pt}
K_{-1} = \left(
  \begin{array}{cc}
    -1 & 4 \\
    4 & -1 \\
  \end{array}
\right)
\label{eq:33}
\end{equation}
Note that the matrices (\ref{eq:33}) have negative determinants, and hence, in contrast to the matrices (\ref{eq:25}),
they describe non-chiral states. We summarize the values of scaling dimensions of excitations in models (\ref{eq:25})
and (\ref{eq:33}) in Table \ref{table}.
%\vskip 2mm
\begin{table}[h]
\begin{tabular}{|c|c|c|}
  \hline
  \hline
  \rule{0pt}{14pt}$K_a$ & $\Delta(\mathbf{n})$ & $\Delta_0, \Delta_{\frac{1}{3}}, \Delta_{\frac{2}{3}},
  \Delta_1, \Delta_{\rm el}$ \\[4pt] \hline\hline
  \rule{0pt}{15pt}$K_5$
   & $\frac{1}{21}(5(n_1^2+n_2^2)+4n_1n_2)$ & $\frac{6}{21},\hspace{5pt}\frac{5}{21},\hspace{5pt}\frac{2}{3},
   \hspace{5pt}\frac{11}{7},\hspace{5pt}5$ \\[7pt] \hline
   \rule{0pt}{15pt}$ K_3 $
  & $\frac{1}{3}(n_1^2+n_2^2)$ & $\frac{2}{3},\hspace{6pt}\frac{1}{3},\hspace{6pt}\frac{2}{3},
  \hspace{6pt}\frac{5}{3},\hspace{6pt}3$ \\[7pt] \hline
   \rule{0pt}{15pt}$K_1$
   & $\frac{2}{3}(n_1^2+n_2^2-n_1n_2)$ & $2,\hspace{6pt}\frac{2}{3},\hspace{6pt}\frac{2}{3},
   \hspace{6pt}2,\hspace{6pt}2$ \\[7pt]  \hline
   \rule{0pt}{15pt}$K_{-1}$
   & $\frac{2}{15}(2(n_1^2+n_2^2)+n_1n_2)$ & $\frac{6}{15},\hspace{5pt}\frac{4}{15},\hspace{5pt}\frac{2}{3},
   \hspace{5pt}\frac{8}{5},\hspace{5pt}4$ \\[7pt]
  \hline
  \hline
\end{tabular}
\caption{Scaling dimensions of excitations in different models of the $\nu =2/3$ state. For each model described
by a matrix $K_a$, we provide the general expression for the scaling dimensions $\Delta(\mathbf{n})$ of quasi-particle operators
labeled by pairs of integer numbers $(n_1,n_2)$. The minimal values $\Delta_q$ for excitations of charge $q$, as well as the
scaling dimensions of electron operators are listed in the right column.}
\label{table}
\end{table}

It is important to note that, for every model, the minimal scaling dimension is
$\Delta_{\rm min} = \Delta_{1/3}$, i.e., the operator of the Laughlin
quasi-particle is the most relevant one. We note that between four models, the
model $K_1$ is presumably most stable with respect to disorder, because it has
the largest scaling dimension $\Delta_0$. Moreover, the electron operator in this
model is the most relevant operator among operators with unit charge. In addition,
numerical simulations \cite{num} and some microscopic considerations \cite{wen23}
confirm that the model with matrix $K_1$ is most likely to describe the $\nu =2/3$ state.
However, some signs of a phase transition in the $\nu=2/3$ state have been observed.\cite{tilted}
This indicates that other models  may also be realized
under certain conditions; see Ref.\ [\onlinecite{Fr-abel}].

\section{The role of Coulomb interactions}
\label{coulomb}

We have shown in Sec.\ \ref{scaling-dimensions} (see also Appendix \ref{correlator}) that, in the non-chiral case, the scaling dimensions
of excitations depend not only on the ``kinematic'' structure of the theory encoded in the $K$-matrix, but also on the angle $\vartheta_+$.
This angle parametrizes the relation between the propagating modes and the electron operators. There is, however, an important class
of systems in which this parameter appears to be uniquely and universally fixed. This is, for instance, the case in a system with two edge
modes and strong Coulomb interactions. This fact has been discussed in Ref.\ [\onlinecite{our}]. The results of the analysis in Ref.\
[\onlinecite{our}] are essentially in perfect agreement with the experimental data of  Refs.\ [\onlinecite{Heiblum2,Basel,Glattli1,mz7}].
Although in Ref.\ [\onlinecite{our}] only the case $\nu=2$ is considered, we will show below that the conclusion of this analysis applies
without significant changes to fractional fluids, too.

Let us assume that effects of disorder are negligible. This may be a reasonable assumption for an electronic MZ interferometer, the size
of which is typically only a few microns. In this case, the generic form of the Hamiltonian is given by a sum of the free Hamiltonian,
the Coulomb interaction term, and a term describing the interaction with an external electromagnetic field $a_\mu$:
\begin{equation}
\mathcal{H} = \mathcal{H}_0 + \mathcal{H}_{\rm C} + \mathcal{H}_{\rm int}[\mathbf{a}].
\end{equation}
%In general, the free Hamiltonian may be written as
%\begin{equation}
%\mathcal{H}_0 = \frac{1}{4\pi}\sum_{i,j} h_{ij}\int dx\, \partial_x\phi_i\partial_x\phi_j,
%\label{eq:11}
%\end{equation}
%where $h_{ij}$ are some unknown constants.
We show below that the actual form of the free Hamiltonian $\mathcal{H}_0$ is not important.

\begin{figure}
\begin{center}
\vskip -3mm \epsfig{width=8cm,height=4cm,figure=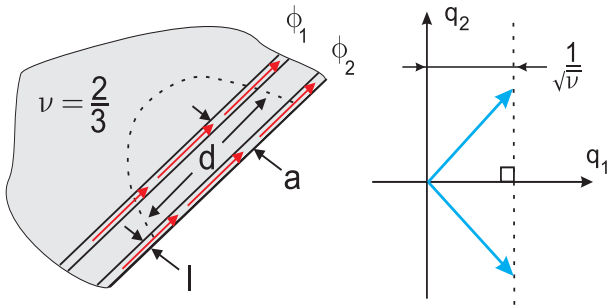}
\caption{(Color online) Illustration of the effects of the strong long-range Coulomb interaction. {\em Left panel:}
important spatial scales at the QH edge are shown: the width of the channels  $l$, the distance between
two channels $a$, and the screening length of the Coulomb interaction $d$. The universal limit is achieved
when $d\gg a,l$. {\em Right panel:} possible configuration of electronic excitations (blue vectors)
in the universal strong interaction limit. Exactly this situation arises at $\nu=2$, as shown in Ref.\ [\onlinecite{our}].}
\label{scales}
\end{center}
\vskip -6mm
\end{figure}

Assuming the distance, $a$,  between the edge channels to be of the order of their thickness, $l$, or smaller (see Fig.\ \ref{scales}
for notations), the Coulomb interaction term can be written as:
\begin{equation}
\mathcal{H}_{\rm C} = (1/2)\int dxdx' \rho_{\rm em}(x)U_C(x-x')\rho_{\rm em}(x'),
\label{eq:43}
\end{equation}
where $\rho_{\rm em}(x)$ is the total one-dimensional charge density at the point $x$, and $U_C(x-x')$ is the Coulomb potential.
We further assume that the interaction is screened at distances $d$, with $L\gg d\gg a$, where $L$ is the size of the interferometer.
This screening can occur due to the presence of the back gate, or the massive air bridge (see Ref.\ [\onlinecite{our}] for a more detailed
discussion). As a consequence, we can neglect the dispersion of the Coulomb interaction and write $U_C(x-y) = V\delta(x-y)$, where the
interaction constant, $V\sim\ln(d/a)$, is large. Finally, the interaction
with an external electromagnetic field is described by
\begin{equation}
\mathcal{H}_{\rm int}[\mathbf{a}] = -\int dx \rho_{\rm em}(x)a_t(x),
\end{equation}
in the gauge $a_x=0$.

In the limit when $\ln(d/a) \gg 1$, the Coulomb interaction exceeds the correlation energy.
One of the most important consequences of this fact is that, independently of the form of free Hamiltonian,
the full Hamiltonian is diagonal in the basis where one mode, $\phi_1$, is charged,  with $Q_1/2\pi\partial_x\phi_1 = \rho_{\rm em}$,
and the other one, $\phi_2$, is a dipole mode of total charge zero. Thus we can write
\begin{equation}
\mathcal{H} = \frac{1}{4\pi}\sum_i\int v_i(\partial_x\phi_i)^2 - \frac{1}{2\pi}Q_1\int a_{t} \partial_{x}\phi_1,
\label{ham-fix}
\end{equation}
where the speed of the charged mode, $v_1 = \sigma_HV$, is much larger than the speed $v_2$ of the dipole mode
determined by the free Hamiltonian (see the discussion at the end of Sec.\ \ref{hydrodynamics}). Comparing
Eq.\ (\ref{ham-fix}) with  (\ref{act-2}), we conclude that $Q_2 = 0$, which means that the dipole mode does not couple
to the external electromagnetic field. The condition of the anomaly cancellation (\ref{canc-2}) therefore implies that
$Q_1 = \sqrt{\nu}$. Thus, the angle between the vector $Q$ and the $q_1$-axis is fixed to the universal value $\vartheta_+ = 0$.
This is illustrated graphically in Fig.\ \ref{scales}.

\section{Experimental determination of charges and scaling dimensions of quasi-particles}
\label{mz-sect}

We have shown in Sec.\ \ref{sec:simplest-two-field} that, for the filling factor $\nu=2/3$, there are several possible models
satisfying all the physical requirements formulated in Sec.\ \ref{bulk-edge}. It is worth noticing that all these models have the
same minimal fractional charge, $1/3$, but different spectra of scaling dimensions. The experiment proposed in this section may
allow one to determine scaling dimensions and, as a result, to identify the physically relevant model of the QH edge.
This experiment is based on the idea to make use of an electronic MZ interferometer.

\begin{figure}[h]
\begin{center}
\vskip -3mm
 \epsfig{width=7.5cm,height=5cm,figure=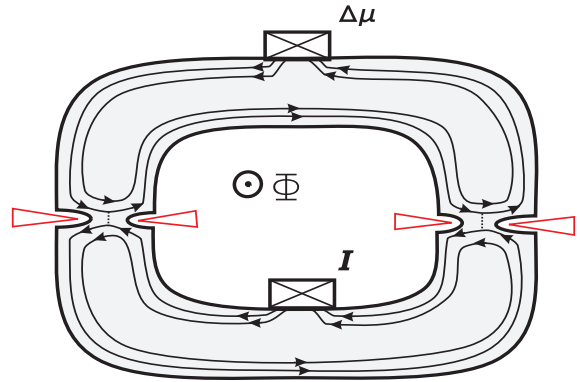}
\caption{(Color online) A Mach-Zehnder interferometer is schematically shown as a Corbino disk, containing a two-dimensional electron gas,
shown in gray shadow. In a strong magnetic field, at a filling factor $\nu=2/3$, two 1D chiral channels are formed at the edges
and propagate along the boundaries of the two-dimensional electron gas (shown by thin black lines). Both channels are partially
transmitted at the left and right QPCs. A bias voltage $\Delta\mu$, applied at the upper Ohmic contact, causes a current $I$ to
flow to the lower Ohmic contact. This current is caused by scattering of quasi-particles at the QPCs and involves an interference
contribution sensitive to the magnetic flux $\Phi$ that can be changed by a slight modulation of
the length of one of the arms.}
\label{schem}
\end{center}
\vskip -6mm
\end{figure}

Electronic MZ interferometers have been realized and investigated experimentally in
Refs.\ [\onlinecite{Heiblum2,Basel,Glattli1,mz7}]. The experimental sample consists
of a two-dimensional electron gas confined to a region of the shape of a so called
{\em Corbino disk} (see Fig.\ \ref{schem}). In the QHE regime, several effectively
one-dimensional conducting channels are formed at the edge. The modes in these edge
channels are used as beams in the electronic MZ interferometer, while two QPCs serve
as beam splitters. Two Ohmic contacts connected to the Corbino disk emit and absorb
electrons. One contact is biased with a voltage $\triangle\mu>0$,
and the other one is grounded and serves as a sink for a current $I$.

There are two paths for quasi-particles to travel from the upper Ohmic contact to the lower one. The first possibilities are to pass
the left QPC and to be reflected off the right QPC. The second possibilities are to bounce off the left QPC and then to pass the right one.
It is easy to see that a nonzero magnetic flux is enclosed by these two paths. Consequently, the current $I$ oscillates as a function
of the magnetic flux through the interferometer. The AB flux may be varied with the help of a modulation gate near one of the arms of
the interferometer that can slightly change the length of this arm (see also the discussion in Appendix \ref{flux-period}).

\begin{figure}[h]
\epsfxsize=4cm
\begin{center}
\epsfbox{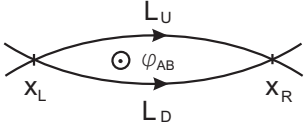}
\end{center}
\caption{Schematic representation of the MZ interferometer. Quasi-particles of electric charge $Q_{\rm em}(\mathbf{n})$
tunnel at points $x_L$ and $x_R$, with tunneling amplitudes $t_{L,\mathbf{n}}$ and $t_{R,\mathbf{n}}$, respectively.
They propagate along paths of length $L_U$ and $L_D$ and acquire an Aharonov-Bohm phase
$\varphi_{AB} =  2\pi Q_{\rm em}(\mathbf{n})\Phi/\Phi_0$. The upper arm is biased with $\Delta\mu$.}
\vspace{-2mm} \label{cnew11}
\end{figure}

We assume that there are several types of excitations, labeled by integers $n_\alpha$, which can tunnel between the arms at the QPCs.
They are created by operators
\begin{equation}
\psi_{\mathbf{n}} = \exp\Big(i\sum_j\! p_j(\mathbf{n})\phi_j\Big),\quad p_j(\mathbf{n}) = \sigma_j\sum_\alpha q^{-1}_{\alpha j}n_\alpha.
\end{equation}
Thus, the tunneling Hamiltonian is given by
\begin{multline}
\mathcal{H}_T = \sum_{\ell,\mathbf{n}}t_{\ell,\mathbf{n}} \psi_{U,\mathbf{n}}^\dag(x_\ell)\psi_{D,\mathbf{n}}(x_\ell)
+ {\rm h.c.} \equiv \\ \equiv \sum_{\ell,\mathbf{n}}\left(A_{\ell,\mathbf{n}}+A^\dag_{\ell,\mathbf{n}}\right),
\label{eq:38}
\end{multline}
where the subscripts $U,D$ indicate that the quasi-particles are created and annihilated at the upper arm and at the lower arm of
the interferometer (see Fig.\ \ref{cnew11}), i.e., at the outer edge and at the inner edge of the Corbino disk.
Moreover, $t_{\ell,\mathbf{n}}$ are the tunneling amplitudes of particles of type ${\bf n}$ at the left and right QPCs, $\ell = L,R$.
These amplitudes include the AB phase shift:
\begin{equation}
\arg \frac{t_{R,\mathbf{n}}}{t_{L,\mathbf{n}}} = 2\pi i Q_{\rm em}(\mathbf{n})\frac{\Phi}{\Phi_0},
\label{phas-ham}
\end{equation}
where $\Phi$ is the flux through the interferometer and $\Phi_0 = hc/e$ is the flux quantum.
Our choice of
tunneling Hamiltonian requires justification, which is presented and discussed
in detail in Appendix \ref{choice}.

The current through the MZ interferometer is defined as a rate of change of the electromagnetic charge
$Q_{\rm em} = \sum_i(Q_i/2\pi)\int dx \partial_x\phi_i$ in one of the arms of the interferometer
(see Fig.\ \ref{cnew11} for notations):
\begin{equation}
\hat{I} = i[\mathcal{H},Q_{\rm em}]=i[\mathcal{H}_T,Q_{\rm em}].
\label{cur}
\end{equation}
Calculating the commutator in Eq.\ (\ref{cur}) with $\mathcal{H}_T$ as in Eq.\ (\ref{eq:38}), we arrive at
the following expression for the current operator:
\begin{equation}
\hat{I} = \sum_{\ell,\mathbf{n}}iQ_{\rm em}(\mathbf{n})(A_{\ell,\mathbf{n}}-A_{\ell,\mathbf{n}}^\dag).
\end{equation}
We evaluate the average current, $I = Tr(\hat{\rho}\hat{I})$, to leading order in the tunneling amplitudes $t_\ell$
\begin{equation}
I = \sum_{\ell,\ell',\mathbf{n}}Q_{\rm em}(\mathbf{n})\int_{-\infty}^{+\infty} dt
\langle[A_{\ell,\mathbf{n}}^\dag(t),A_{\ell',\mathbf{n}}(0)] \rangle,
\label{eq:41}
\end{equation}
where the operators $A_{\ell,\mathbf{n}}^\dag$, $A_{\ell,\mathbf{n}}$ are taken in the interaction representation,
and averaging is defined as $\langle\dots\rangle := Tr\hat{\rho}_0(\dots)$, where $\hat{\rho}_0$ is the density
operator of disconnected arms. In Eq.\ (\ref{eq:41}) we have taken into account that
$\langle \psi^\dag_{\mathbf{n}}\psi_{\mathbf{m}}\rangle \propto \delta_{\mathbf{n},\mathbf{m}}$,
which is a consequence of zero-modes.

It is easy to see that expression (\ref{eq:41}) for the current is a sum of four terms:
$I = \sum_{\ell\ell'}I_{\ell\ell'}$, where $\ell,\ell' = L,R$. The first two terms,
\begin{equation}
I_{\ell\ell} = \sum_{\mathbf{n}}Q_{\rm em}(\mathbf{n})\int_{-\infty}^{+\infty} dt
\langle[A_{\ell,{\mathbf{n}}}^\dagger(t),A_{\ell,{\mathbf{n}}}(0)]
\rangle,
\end{equation}
correspond to incoherent tunneling at either one of the two QPCs. The other two terms depend on the magnetic
flux $\Phi$ and lead to interference:
\begin{multline}
I_{\Phi} \equiv I_{LR}+I_{RL}  \\ = 2\sum_{\mathbf{n}}Q_{\rm em}(\mathbf{n}){\rm Re}\;\int_{-\infty}^{+\infty}
dt \langle[A_{R,{n}}^\dagger(t),A_{L,{n}}(0)] \rangle.
\end{multline}
We focus our attention on the interference term, because it allows us to discriminate between contributions from
different excitations. Using Eq.\ (\ref{eq:38}),
%$\triangle\mu >0$
we write:
\begin{multline}
\label{interf}
I_{\Phi} = 2\sum_{\mathbf{n}}Q_{\rm em}(\mathbf{n}){\rm Re}\;t_{L,\mathbf{n}}t^*_{R,\mathbf{n}}
\int_{-\infty}^{+\infty} dt \\
\left\{
\langle \psi_{D,\mathbf{n}}(x_R,t)\psi_{D,\mathbf{n}}^\dagger(x_L,0)\rangle
\langle \psi_{U,\mathbf{n}}^\dagger(x_R,t)\psi_{U,\mathbf{n}}(x_L,0)\rangle
\right.\\
\left.
-\langle \psi_{D,\mathbf{n}}^\dagger(x_L,0)\psi_{D,\mathbf{n}}(x_R,t)\rangle
\langle \psi_{U,\mathbf{n}}(x_L,0)\psi_{U,\mathbf{n}}^\dagger(x_R,t)\rangle
\right\}.
\end{multline}
The correlation functions are evaluated in Appendix \ref{corr-f-calc}. The result is:
\begin{multline}
i\langle\psi_{\mathbf{n}}^\dagger(x,t)\psi_{\mathbf{n}}(0,0)\rangle \propto \exp [i\varphi_0(\mathbf{n})]
\\ \times
\prod_i\left\{\frac{v_i}{\pi T}\sinh[\pi T(t+\sigma_i\frac{x}{v_i})]\right\}^{-\delta_i(\mathbf{n})}.
\label{step6}
\end{multline}

The phase $\varphi_0(\mathbf{n})$ in this equation is determined by the structure of zero-modes
and requires a separate consideration. Introducing zero-modes $\phi_i$ and $\pi_i$ via
$\phi_i(x) = \phi_i + 2\pi x\pi_i+osc.$, we write the corresponding term in Eq.\ (\ref{ham-fix})
as $\mathcal{H}_0 = \pi W\sum_i v_i\pi_i^2$, where $W$ is the total size of the system.
The total charge at an edge is given by
$Q_{\rm em} = W\langle\pi_1\rangle$. The expectation values of the zero-modes can be related to the
applied voltage bias $\Delta\mu$ by appealing to the well known electrostatic formula
$\Delta\mu = \delta\langle\mathcal{H}_0\rangle/\delta Q_{\rm em}$. From this equation it follows
that $\langle\pi_1\rangle = \Delta\mu/v_1$, while for the dipole mode $\langle\pi_2\rangle =0$,
because it is not biased. \cite{footnote5aa} We assume that charge fluctuations are negligible due to the
large capacitances of edge channels connected to Ohmic contacts. Thus, the contribution of zero-modes
is given by $\langle e^{2\pi ip_j\pi_j(x+\sigma_jv_j t)}\rangle = e^{2\pi ip_j\langle\pi_j\rangle(x+\sigma_jv_jt)}$.
Substituting the expectation values of zero modes, we find the phase of correlation function (\ref{step6}):
\begin{multline}
\varphi_0(\mathbf{n}) \equiv \sum_i p_i\langle\pi_i\rangle(x+\sigma_iv_it)\\ = \Delta\mu(t+\sigma_i\frac{x}{v_1})
\sum_\alpha q^{-1}_{\alpha 1} n_\alpha.
\label{phi0}
\end{multline}
Apparently, this phase is linear in the bias $\Delta\mu$.

In the zero-temperature limit, $T=0$, the correlation functions are given by formula (\ref{corr-m}).
The time integral in Eq.\ (\ref{interf}), at small biases $\Delta\mu\ll v_i/L_\alpha$, $\alpha = U,D$, then yields
\begin{equation}
I_{\Phi} = \sum_{\mathbf{n}} C_\mathbf{n}(\Delta\mu)^{2\Delta(\mathbf{n})-1}
\cos\left[ 2\pi Q_{\rm em}(\mathbf{n})\frac{\Phi}{\Phi_0}\right],
\label{curr-v}
\end{equation}
where the $C_\mathbf{n}$ are some (unimportant) constants. In the high-temperature limit,
correlation function (\ref{step6}) scales as
\begin{equation}
i\langle\psi_{\mathbf{n}}^\dagger(x,t)\psi_{\mathbf{n}}(0,0)\rangle
\propto \exp\big[-\sum_i \pi T\delta_i(\mathbf{n}) |t+\sigma_i\frac{x}{v_i}|\big].
\label{step7}
\end{equation}
The oscillating part of the current at small bias then takes the following form
(see Appendix \ref{corr-f-calc}):
\begin{multline}
\frac{I_{\Phi}}{ \Delta\mu} = \sum_{\mathbf{n}} C'_\mathbf{n}T^{2\Delta(\mathbf{n})-1}
e^{-\pi T/T_0(\mathbf{n})}\cos\left[ 2\pi Q_{\rm em}(\mathbf{n})\frac{\Phi}{\Phi_0}\right],
\label{curr-t}
\end{multline}
where the $C'_\mathbf{n}$ are constants, and the characteristic energy scale is given by
\begin{equation}
\frac{1}{T_0(\mathbf{n})} = \min_{\alpha',i'}\sum_{\alpha, i} \delta_i(\mathbf{n})
|\frac{\sigma_iL_\alpha}{v_i}-\frac{\sigma_{i'}L_{\alpha'}}{v_{i'}}|.
\label{energ}
\end{equation}
For a symmetric interferometer, in the limit $v_1 \gg v_2$, this expression simplifies to
\begin{equation}
T_0^{-1}(\mathbf{n}) = \min(\delta_1(\mathbf{n}),\delta_2(\mathbf{n}))\frac{L}{v_2}.
\end{equation}

The range of applicability of our result (\ref{curr-v}) is limited by the conditions
$\Delta\mu\ll v_i/L_\alpha$, $\alpha = U,D$. Outside of this range, the dependence
of the visibility on the bias is non-monotonic, because of charging effects,
as has been observed in the experiments [\onlinecite{Heiblum2}-\onlinecite{mz7}].
Moreover, the behavior (\ref{curr-v}) is valid only if $\Delta\mu>T$.
For typical experiments \cite{Heiblum2,Basel,Glattli1,mz7} this implies that $1 \mu V < \Delta\mu < 10 \mu V$.
We conclude that it would not be easy, but possible, in principle, to extract the exponents of
the power-law behavior in Eq.\ (\ref{curr-v}).

To summarize, in contrast to theoretical works where ad-hoc Klein factors are used, \cite{Feldman}
we predict periods of AB oscillations larger than $\Phi_0$; (for a discussion of Klein factors see Appendix \ref{choice}).
The easiest way to experimentally detect larger periods is to compare periodicities in the weak tunneling and in the weak
backscattering regimes. Eqs.\ (\ref{curr-v}) and (\ref{curr-t}) are the central results of our paper.
They can be used to discriminate between different effective models. Namely, they can be fitted by measuring the current
through an MZ interferometer as a function of the magnetic flux $\Phi$ and of the bias $\Delta\mu$. Evaluating the Fourier
transform with respect to $\Phi$, one can investigate the scaling in $\Delta\mu$ of different harmonics corresponding to
contributions of the most relevant excitations for any charge $Q_{\rm em}$. To identify the correct model, one must
compare the experimentally measured scaling dimensions $\Delta$ with those in the table of Sec.\ \ref{sec:simplest-two-field}.
Once the correct model (i.e., its $K$-matrix) is identified, one may use the temperature dependence (\ref{curr-t}) as an
independent check of the theory.

\section{Conclusion}

In the last decade, several proposals for experimental tests of the physics at a QH edge have been made.
They are based on measurements of the electric charge, the statistical phases and the scaling dimensions
of quasi-particles. Some of them have been realized and have shed light on the properties of fractional
QH edges. However, some experiments have brought up open questions. For instance, in the experiment
[\onlinecite{Gr-Chang}], the I-V curve has shown a perfect power-law behavior. However, the measured
exponents, which are thought to be proportional to the scaling dimensions of the quasi-particle operators,
have turned out to be different from those predicted by theory. Thus, MZ interferometers, which have already
shown several interesting features, may be considered to be promising tools for probing the properties of
the QH edge.

In this paper, we have reviewed the construction of a low-energy theory\cite{Fr-abel,Fr-non-abel} of fractional QH edges
based on anomaly cancellation. We have shown that for $\nu = 1/m$, where $m$ is an odd integer, it can be described by a hydrodynamical
model, while other filing factors require the introduction of several edge channels (\ref{act-2}).
%The classification of such models, based on the physical requirements formulated at the beginning of Sec.\ \ref{bulk-edge},
%amounts to the classification of matrices (\ref{phase-cond}) for statistical phases of electrons.
Quasi-particle operators in each model are found to be indexed by vectors in the dual of an odd, integral lattice. \cite{Fr-abel}
Their charges and statistical phases are given by Eqs.\ (\ref{phas-v}) and (\ref{charg-v}). We have illustrated the
classification of effective models with the example of fluids with filling fraction $\nu = 2/m$, and, in particular, with
$\nu = 2/3$. We have shown that, for $\nu = 2/3$, there are at least four inequivalent models satisfying all physical conditions
and having the smallest possible number of fields. It is important to note that, in every effective model, the minimal
fractional charge is $1/3$.

For models with two fields, we have shown that Coulomb interactions lead to universal values of electromagnetic couplings.
This universality allowed us to evaluate the scaling dimensions of quasi-particles, see (\ref{scaling-d}), with the result given in
Eq.\ (\ref{non-ch}). We have calculated the AB-oscillating contribution to the current through an MZ interferometer at low and high
temperatures, Eqs.~(\ref{curr-v}) and (\ref{curr-t}), and shown that the Fourier spectrum of the current as
a function of the flux can be used to extract the scaling dimensions of quasi-particle operators. This, in turn, leads to the
possibility to discriminate between different effective models.

Our method to identify the correct model can be applied to fluids with arbitrary filling fractions and can be summarized as follows:

\begin{itemize}

\item First, for a given filling fraction $\nu$, one should find solutions of Eq.\ (\ref{cond-main}) for $K$-matrices,
up to equivalence, as described in Eq.\ (\ref{equiv}). In other words, one must identify the effective models satisfying the
physical requirements formulated in Sec.\ \ref{bulk-edge}.
The most interesting solutions are those with the smallest possible number of fields and minimal statistical phases of
electron field operators.

\item Second, using Eqs.\ (\ref{charg-v}) and (\ref{scaling-d}), one should calculate the spectra of charges and scaling
dimensions for every model.

\item Finally, one should attempt to measure the $\Delta\mu$-scaling of the Fourier components of the current trough an
MZ interferometer and compare it with theoretical predictions, in order to identify a correct model.

\end{itemize}

An important aspect of our theory is that it predicts AB oscillations with quasi-particle periodicity in the
gate modulated magnetic flux, i.e.,
with periods equal to several electronic periods. This periodicity allows one to separate the contributions of different excitations
to the  current.
%However, it appears only as a function of the voltage bias on the modulation gate that deforms the AB path,
%while, for a point-like flux, the standard periodicity is restored.
In the context of our theory, the quasi-particle periodicity
is related to our choice of a tunneling Hamiltonian, which is different from the one in Refs.\ [\onlinecite{Feldman}] and
[\onlinecite{Averin}] and leads to a non-commutativity of tunneling Hamiltonians at different spatial points. This non-commutativity
originates from the topological character of quasi-particle excitations in a fractional QH state and is a consequence of open boundary
conditions specific to the MZ interferometer.

We think that the non-commutativity of tunneling Hamiltonians calls for additional theoretical analysis and, possibly,
for experimental tests. A theoretical analysis of this problem should include a concrete model of Ohmic contacts, which
may influence the physics of processes in an MZ interferometer. It is also interesting to generalize our analysis to
fractions, such as $\nu=5/2$, which are possibly described by non-Abelian QH states, and to the case of FP interferometers,
where new physics may emerge.

\begin{acknowledgments}
We thank V.\ Cheianov and O.\ Ruchayskiy for valuable discussions.
This work has been supported by the Swiss National Foundation.
\end{acknowledgments}

\appendix

\section{Calculation of scaling dimensions.\label{correlator}}

Using expression~(\ref{eq:23}) for $\delta_i$, we calculate the total scaling dimension $\Delta$.
As it has already been mentioned, in the general (non-chiral) case, the scaling dimension is a function
of the full matrix $q$. Therefore, apart from the matrix $K$, it depends only on one additional variable,
which can be fixed by choosing the Hamiltonian. We are interested in a matrix $K$ of the following form:
\begin{equation}
K = \left(
          \begin{array}{cc}
            a & b \\
            b & a \\
          \end{array}
        \right).
\end{equation}
The connection between matrices $K$ and $q$ in the non-chiral case is $K = q\sigma q^T$.
Therefore we can introduce the following parametrization of the matrix $q$,
\begin{equation}
q/\sqrt{a} = \left(
  \begin{array}{cc}
    \cosh\vartheta_1 & \sinh\vartheta_1 \\
    \cosh\vartheta_2 & \sinh\vartheta_2 \\
  \end{array}
\right),
\end{equation}
where $\cosh (\vartheta_2-\vartheta_1) = b/a$. Although the case $a<0$ requires a different parametrization,
it leads to the same result.

The evaluation of the scaling dimension $\Delta(\mathbf{n}) = \mathbf{n}(qq^T)^{-1}\mathbf{n}$ requires to invert the matrix:
\begin{equation}
qq^T = a\left(
             \begin{array}{cc}
               \cosh 2\vartheta_1 & \cosh (\vartheta_1+\vartheta_2) \\
               \cosh (\vartheta_1+\vartheta_2) & \cosh 2\vartheta_2 \\
             \end{array}
           \right).
\label{qqt}
\end{equation}
For convenience, we introduce the angle $\vartheta_+ = \vartheta_1 +\vartheta_2$ which takes arbitrary values,
and the angle $\vartheta_- = \vartheta_1 -\vartheta_2$ which is fixed by the condition
\begin{equation}
\label{eq:42}
\cosh\vartheta_- = b/a.
\end{equation}
Inverting the matrix $qq^T$ (see Eq.\ (\ref{qqt})), we find the following expression for the scaling dimensions:
\begin{equation}
\Delta(\mathbf{n}) = \frac{1}{b^2-a^2}\left(
                                \begin{array}{c}
                                  n_1 \\
                                  n_2 \\
                                \end{array}
                              \right)^T \left(
             \begin{array}{cc}
                A_- & B \\
                B & A_+ \\
             \end{array}
           \right) \left(
                     \begin{array}{c}
                       n_1 \\
                       n_2 \\
                     \end{array}
                   \right),
\end{equation}
where
\begin{eqnarray*}
A_\pm  &=& b\cosh\vartheta_+ \pm\sqrt{b^2-a^2}\sinh\vartheta_+,
\\ B &=& -a\cosh\vartheta_+.
\end{eqnarray*}
We see that $\Delta$ indeed depends on the additional free parameter, the angle $\vartheta_+$.

If we assume that the strong long-range Coulomb interaction is a dominant contribution to the Hamiltonian, then we may approximate $q_{11} = q_{12}$, or equivalently, $\vartheta_2 = -\vartheta_1$.
This condition leads to $\cosh\vartheta_+ = 1$ and $\sinh\vartheta_+ = 0$, so that the expression for the scaling dimension simplifies:
\begin{equation}
\Delta(\mathbf{n}) = \frac{1}{b^2-a^2}\left(
                                \begin{array}{c}
                                  n_1 \\
                                  n_2 \\
                                \end{array}
                              \right)^T \left(
             \begin{array}{cc}
               b & -a \\
               -a & b \\
             \end{array}
           \right) \left(
                     \begin{array}{c}
                       n_1 \\
                       n_2 \\
                     \end{array}
                   \right).
\end{equation}
Calculating the product, we arrive at the final result (\ref{non-ch}).

Next, we evaluate the exponents $\delta_1$ and $\delta_2$. In the chiral case,
which we consider as an example, $K = qq^T$, and the following parametrization is required
\begin{equation}
q/\sqrt{a} = \left(
  \begin{array}{cc}
    \cos\vartheta_1 & \sin\vartheta_1 \\
    \cos\vartheta_2 & \sin\vartheta_2 \\
  \end{array}
\right)
\end{equation}
with the condition that $\cos (\vartheta_2-\vartheta_1) = b/a$.
Then, using definition (\ref{eq:23}), we find
\begin{eqnarray}
\delta_1(\mathbf{n}) = \frac{a}{a^2-b^2}(n_1\cos\vartheta_1-n_2\cos\vartheta_2)^2,
\\ \delta_2(\mathbf{n}) = \frac{a}{a^2-b^2}(n_1\sin\vartheta_1-n_2\sin\vartheta_2)^2.
\end{eqnarray}
Thus, we see that by measuring the exponents $\delta_1$ and $\delta_2$ one can in principle
extract the parameter $\vartheta_+$. However, we stress again that, for strong Coulomb
interaction, $\vartheta_+ = 0$. In this case, expressions for exponents simplify.
Namely, taking into account that $a+b = 2/\nu$, we find that
\begin{equation}
\delta_1(\mathbf{n}) = \frac{(n_1-n_2)^2}{2(a-b)},\quad \delta_2(\mathbf{n}) = \frac{\nu}{4}(n_1+n_2)^2.
\end{equation}
In the non-chiral case, analogous calculations lead to similar expressions
\begin{equation}
\delta_1(\mathbf{n}) = \frac{(n_1-n_2)^2}{2(b-a)},\quad \delta_2(\mathbf{n}) = \frac{\nu}{4}(n_1+n_2)^2.
\end{equation}

\section{Correlation function at finite temperature and asymptotics of tunneling current}
\label{corr-f-calc}

%After we have introduced the model in Sec.\ \ref{sec:multi-field-edge}, the derivation of the
%correlation function of local excitations (\ref{exc}) is relatively simple.
Using the Gaussian character of the edge fields $\phi_i$, the correlation functions
for the operators $\psi_{\bf n} = \exp\big(i\sum_jp_j({\bf n})\phi_j\big)$ may be written in the following form:
\begin{equation}
i\langle\psi_{\bf n}^\dagger(x,t)\psi_{\bf n}(0,0)\rangle
= e^{i\varphi_0}K_{\bf n}(x,t),
\label{step1}
\end{equation}
where the first factor is the zero-mode contribution given by Eq.\ (\ref{phi0}), while the
function $K_{\bf n}$ is the fluctuation part:
\begin{equation}
\ln[K_{\bf n}(x,t)]= \sum_{ij}p_i({\bf n})p_j({\bf n}) \langle[\phi_{i}(x,t)-\phi_{i}(0,0)]\phi_{j}(0,0)\rangle.
\label{step2}
\end{equation}
Introducing the notation $X_j\equiv x+\sigma_jv_j t$, we express fields in terms of creation and annihilation operators,
\begin{equation}
\phi_{j}(x,t) = i\sum_{k}\sqrt{\frac{2\pi}{Wk}}\big[
a_{j}(k)e^{ikX_j} + a^\dagger_{j}(k)e^{-ikX_j}\big],
\label{step3}
\end{equation}
where $W$ is the system size.
Substituting this expression into Eq.\ (\ref{step2}), we obtain
\begin{eqnarray}
\ln[K_{\bf n}] = \sum_j p_j^2({\bf n})\int_0^{\Lambda}
\frac{dk}{k}\left\{f_j(k)(e^{-ik X_j}-1)\right.\nonumber\\
+\left.[1+f_j(k)](e^{ik X_j}-1)\right\} ,
\label{step4}
\end{eqnarray}
where $f_j(k)=[\exp(\beta v_j k)-1]^{-1}$ are the boson occupation numbers,
and $\Lambda$ is an ultraviolet cutoff.

The best way to proceed is to expand the occupation numbers in Boltzmann factors,
$f_j(k)=\sum_{m=1}^\infty\exp(-\beta v_j m\,k)$, and integrate each term separately.
This gives
\begin{equation}
\ln[K_{\bf n}] = -\sum_{j}p_j^2({\bf n})\!\!\sum_{m=-\infty}^{\infty}
\ln[\Lambda(i\beta v_j m-X_j)].
\label{step5}
\end{equation}
Combining this expression with Eq.\ (\ref{step1}), we finally arrive at the following
result:
\begin{equation}
i\langle\psi_{\bf n}^\dagger(x,t)\psi_{\bf n}(0,0)\rangle\propto e^{i\varphi_0}
\prod_i\left[\frac{v_i}{\pi T}\sinh\left(\pi \frac{TX_i}{v_i}\right)\right]^{-\delta_i({\bf n})}.
\label{step5a}
\end{equation}
The scaling exponents $\delta_i$ are calculated in Appendix \ref{correlator}.

Next, we use high-temperature limit (\ref{step7}) of the correlation function
(\ref{step5a}) to calculate the high-temperature asymptotics of the tunneling current.
Substituting the correlation function  (\ref{step7}) into Eq.\ (\ref{interf}), we obtain
the following expression:
\begin{equation}
I_\Phi \propto {\bf Re}\int_{-\infty}^{+\infty} dt\, e^{i\Delta\mu t}T^{2\Delta({\bf n})}
e^{-\pi T\sum\limits_{i,\alpha}|t+\sigma_iL_\alpha/v_i|\delta_i({\bf n})}
\end{equation}
In the limit $T\gg\Delta\mu$, we approximate $e^{i\Delta\mu t} \simeq 1 + i\Delta\mu t$,
where only the second term makes a non-zero contribution:
\begin{equation}
I_\Phi \propto \Delta\mu T^{2\Delta({\bf n})}\int_{-\infty}^{+\infty} dt\, t
\cdot e^{-\pi T\sum\limits_{i,\alpha}|t+\sigma_iL_\alpha/v_i|\delta_i({\bf n})}
\end{equation}
In the high-temperature limit, the largest contribution to this integral comes from
a small region around one of the points $t=-\sigma_iL_\alpha/v_i$, where
the argument of the exponential function acquires the smallest absolute value.
Then the time integral can be estimated as
\begin{multline}
\int_{-\infty}^{+\infty} dt\, t
\cdot e^{-\pi T\sum\limits_{i,\alpha}|t+\sigma_iL_\alpha/v_i|\delta_i({\bf n})}
\propto T^{-1}e^{-\pi T/T_0({\bf n})},
\end{multline}
where the energy scale $T_0({\bf n})$ is given by Eq.\ (\ref{energ}).
Using this result, we finally arrive at the asymptotics (\ref{curr-t})
of the oscillating part of the current.

\section{Tunneling Hamiltonian}
\label{choice}

In this appendix, we discuss two important questions concerning the form of tunneling Hamiltonians.
The first question is whether one needs to introduce Klein factors \cite{Safi} to ensure
commutativity of the tunneling Hamiltonians at spatially separated points.
We argue that the correct choice of the tunneling Hamiltonian generally leads to AB-oscillations in the quasi-particle current.
The second question is about the value of the AB phase shift that should be included in tunneling amplitudes.

\subsection{Non-commutativity and Klein factors}
\label{Klein}

For simplicity let us consider the case of filling factor $\nu = 1/m$, where only one channel at each
edge of the MZ interferometer is present. Tunneling
of Laughlin quasi-particles is described by the Hamiltonian $\mathcal{H}_L+\mathcal{H}_R$,where
\begin{equation}
\mathcal{H}_\ell = t_\ell\psi^\dag_{U}(x_\ell)\psi_D(x_\ell) +t^*_\ell\psi_D^\dag(x_\ell)\psi_U(x_\ell),\; \ell = L,R,
\label{ham-0}
\end{equation}
are the contributions at two spatially separated points $x_L$ and $x_R$.
It is interesting to calculate the commutator of $\mathcal{H}_L$ and $\mathcal{H}_R$. For this purpose,
we first have to find
the commutation relations for quasi-particle operators in a system with two edges.

We remind the reader that only local excitations may tunnel at QPCs.
One of the conditions of locality reads:
\begin{equation}
[\partial_x\phi_U,\psi_D] = 0,\quad [\partial_x\phi_D,\psi_U] = 0,
\end{equation}
which means that a quasi-particle at one edge does not create a charge density at the other edge. This condition
implies that $[\psi_U,\psi_D] = 0$. On the other hand, for quasi-particle operators at the same edge,
in the case of open boundary conditions, we have that
\begin{equation}
\psi_\alpha(x)\psi_\alpha(x') = e^{(i\pi/m){\rm sign}(x'-x)}\psi_\alpha(x')\psi_\alpha(x),
\label{commut-0}
\end{equation}
where $\alpha=U,D$, and the coordinate $x$ starts at the Ohmic contact and increases in the
direction of the chirality of the corresponding channel.

The sign of the statistical phase in this expression is determined by the sign of the right-hand-side of Eq.~(\ref{normalization})
and depends on the chirality of the channel.
Assuming $x_R>x_L$ (see Fig.\ \ref{cnew11}), we arrive at the following result:
\begin{multline}
\psi^\dag_{U}(x_L)\psi_D(x_L)\psi^\dag_{U}(x_R)\psi_D(x_R)\\ =
e^{2\pi i/m}\psi^\dag_{U}(x_R)\psi_D(x_R)\psi^\dag_{U}(x_L)\psi_D(x_L).
\label{res-com}
\end{multline}
Thus we conclude that for the tunneling Hamiltonians defined in Eq.\ (\ref{ham-0}), $[\mathcal{H}_L,\mathcal{H}_R] \neq 0$.
Note that this is not the case for a Fabry-Perot type interferometer, where the order of tunneling points with respect to
chirality is different on different arms.
%(see Fig.\ \ref{mz-2}).
Therefore contributions to the statistical phase from inner and outer edges cancel, and the Hamiltonians
$\mathcal{H}_L$ and $\mathcal{H}_R$ commute.

\begin{figure}[hbt]
\epsfxsize=8cm
\begin{center}
\epsfbox{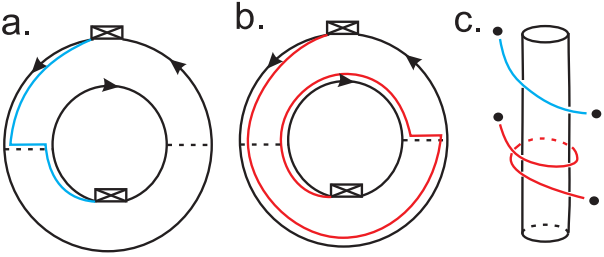}
\end{center}
\caption{(Color online) The Mach-Zehnder interferometer and Wilson lines are schematically shown.
Tunneling Hamiltonians may be expressed in terms of integrals
along the lines between Ohmic contacts, see Eq.~(\ref{wils}). {\em Panel a:}
the Wilson line (drawn in blue) corresponds to tunneling at the left QPC,
described by the Hamiltonian $\mathcal{H}_L$. {\em Panel b:} the Wilson line (drawn in red)
for the Hamiltonian $\mathcal{H}_R$. {\em Panel c:}
``time-expanded'' representation of the product $\mathcal{H}_R\mathcal{H}_L$.
We see that the two lines cannot be ``topologically exchanged'',
which means that the two Hamiltonians $\mathcal{H}_L$ and $\mathcal{H}_R$ do not commute.} \vspace{-2mm}
\label{mz-1}
\end{figure}

\begin{figure}[hbt]
\epsfxsize=8cm
\begin{center}
\epsfbox{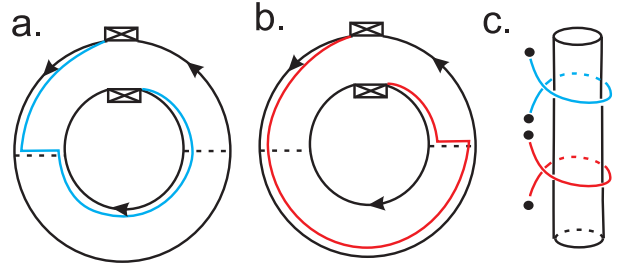}
\end{center}
\caption{(Color online) The Fabry-Perot interferometer and Wilson lines are schematically shown. {\em Panel a:}
Wilson line (drawn in blue), which represents the Hamiltonian $\mathcal{H}_L$, differs from the one
for the MZ interferometer. {\em Panel b:} Wilson line (drawn in red) for the Hamiltonian $\mathcal{H}_R$.
{\em Panel c:} time-expanded representation of two Wilson lines. One can see that the lines are
topologically identical, therefore Hamiltonians $\mathcal{H}_L$ and $\mathcal{H}_R$ commute.} \vspace{-2mm}
\label{mz-2}
\end{figure}

There is a useful geometrical illustration of the commutation relations discussed above.
It is a well known fact about the Chern-Simons effective theory that the quasi-particle operator can
be represented as a Wilson line. \cite{CS-theory}
According to the boundary conditions for the excitations in the MZ interferometer,
we choose the Ohmic contacts as end points of Wilson lines (see the discussion in Appendix \ref{cs-bulk}).
Thus the tunneling Hamiltonians are given by Wilson lines going from one Ohmic contact to the other:
\begin{equation}
\mathcal{H}_\ell = t_\ell \exp\Big[\frac{i}{\sqrt{m}}\int_{\gamma_\ell} dr^\mu
b_\mu \Big] + t^*_\ell \exp\Big[\frac{i}{\sqrt{m}}\int_{-\gamma_\ell} dr^\mu b_\mu \Big],
\label{wils}
\end{equation}
where $b_\mu$ is the Chern-Simons field, and
$\gamma_\ell$ is the line going from the upper Ohmic contact to the lower one
through the $\ell$-th QPC, see Figs.\ \ref{mz-1}a and \ref{mz-1}b.\cite{wline}
The product of these operators, $\mathcal{H}_R\mathcal{H}_L$,
is represented by the configuration of Wilson lines shown schematically in Fig.\ \ref{mz-1}c.
On the other hand, the permutation $\mathcal{H}_L\mathcal{H}_R$
of these operators may be represented by the lines oppositely ordered in time.
According to Chern-Simons theory the corresponding braidings again yield result (\ref{res-com}).
Interestingly, for a FP interferometer, the corresponding braidings are trivial (see Fig.\ \ref{mz-2});
therefore the tunneling Hamiltonians commute.

Non-commutativity of tunneling Hamiltonians has often been claimed in the literature to be unphysical.
It has therefore been proposed to use Klein factors to ensure their commutativity. To clarify the
nature of Klein factors, we consider for a moment the boundary conditions which correspond to
a closed geometry. In this case the quasi-particle operators $\psi_\alpha=e^{(i/\sqrt{m})\phi_\alpha (x)}$
are $m$-fold multi-valued operators. This is because the field $\phi_\alpha (x)$
is in fact an integral of the density $\rho_\alpha$ over a closed contour that can have
different numbers of winding, $\kappa = 0,1,\ldots, m-1$. Hence we can write
\begin{equation}
\phi_{\alpha}(x) = \phi_{\alpha} + 2\pi(x + \kappa W_\alpha)\pi_{\alpha}+ osc., \hspace{5pt} \alpha=U,D,
\end{equation}
where $\phi_\alpha$ and $\pi_\alpha$ are the zero modes, and $W_\alpha$ is the length of the
boundary $\alpha$.

We stress that only {\em relative} branch numbers are physically observable.
Indeed, the two zero modes $\pi_{\alpha}$, where $\alpha=U,D$, are quantized as
$W_\alpha\pi_{\alpha} = N_{\alpha}/\sqrt{m}$.
However, any closed electronic system contains an integer number, $N$, of electrons,
which imposes the following constraint on the total charge operator:
$(W_U\pi_U+W_D\pi_D)/\sqrt{m}=N$. This limitation on the eigenvalues of the operators
$\pi_{\alpha}$, which any physically allowed state should satisfy, leads to the fact that
\begin{equation}
F \equiv e^{(2\pi i/\sqrt{m})W_U\pi_U} = e^{-(2\pi i/\sqrt{m})W_D\pi_D}.
\end{equation}
We use the winding number $\kappa$ as an additional index that denotes the quasi-particle branch,
$\psi_{\alpha,\kappa} = \psi_{\alpha}F^\kappa$. Here $\psi_{\alpha}$ stands for the case
where $\kappa=0$, i.e.,  for the branch that starts from one of the Ohmic contacts.
In the language of Wilson lines, every branch is given by a line with the corresponding number of windings.

If several branches are present, then one should take into account
all possible processes in the tunneling Hamiltonian
including those that change the quasi-particle branch number.
Therefore, in general, the tunneling Hamiltonian can be written as
\begin{equation}
\mathcal{H_\ell} = \sum_{\kappa'} t_{\ell,\kappa'}\psi_{D,\kappa}^\dag(x_\ell)\psi_{U,\kappa+\kappa'}(x_\ell)
+{\rm h.c.}
\label{ham-gen}
\end{equation}
The commutation relations for quasi-particle operators (\ref{commut-0}) are easily generalized:
\begin{multline}
\psi_\kappa(x)\psi_{\kappa'}(x') = e^{(2i\pi/m)(\kappa-\kappa')}\\
\times e^{(i\pi/m){\rm sign}(x'-x)}\psi_{\kappa'}(x')\psi_\kappa(x),
\label{commut-gen}
\end{multline}
and we arrive at the important conclusion that, in general,
Hamiltonians (\ref{ham-gen}), taken at different spatial points, do not commute.

In order to put our discussion into the context of previous work,
we rewrite Hamiltonian (\ref{ham-gen}) in a slightly different form:
\begin{equation}
\mathcal{H_\ell} = \sum_{\kappa} t_{\ell,\kappa}F^\kappa\psi_{D}^\dag(x_\ell)\psi_{U}(x_\ell)+{\rm h.c.},
\label{compare}
\end{equation}
where the operator $F$ introduced earlier obviously plays the role of a Klein factor.
Interestingly, in the specific case when only the amplitudes $t_{L,1}$ and $t_{R,0}$ are non-zero,
tunneling Hamiltonians (\ref{compare}) do commute at different spatial points.
Moreover, in the language of Wilson lines the multiplication of the tunneling Hamiltonian
with the Klein factor $F$ is equivalent to adding a loop to the corresponding Wilson line, so that
it goes all the way around the interferometer. It is easy to see that adding such a loop to the blue line
in Fig.\ \ref{mz-1} makes it topologically equivalent to the red line. Therefore the Hamiltonians,
after such manipulation, indeed commute.

Here we have to admit that the Klein factors introduced earlier in the literature
are usually supposed to commute with the quasi-particle operators $\psi_{\alpha}$
(see, for instance, Refs.\ [\onlinecite{Safi}] and [\onlinecite{Vishvesh}]).
In other words, they act on some additional Hilbert space.
This situation, however, is not satisfactory, because it contradicts the very well known
aspect of the QHE that there exists a gap for excitations in the bulk, and the only degrees of
freedom available are the edge excitations $\phi_\alpha(x)$, including zero modes $\phi_\alpha$ and
$\pi_\alpha$. Therefore, we think that Klein factors should be expressed in terms of the
same modes as the operators $\psi_\alpha$ (see, e.g., Ref. [\onlinecite{klein-true}]).
On the other hand, choosing specific Hamiltonians that commute
does not appear to be physical and needs, to say the least, additional justification.
Moreover, it is important that Klein factors may be introduced only in a system with a closed geometry,
where multivalued excitations may exist. The strong coupling to Ohmic contacts, as in the case of an
MZ interferometer considered in this paper, requires open boundary conditions. Therefore we insist
that our form (\ref{ham-0}) of the tunneling Hamiltonian is correct and use it in Sec.\ \ref{mz-sect}
for calculations.

Our next remark concerns the statement made in the literature \cite{Feldman,Averin,Stern,Feldman2}
that it is impossible to observe a coherent part of the quasi-particle current.
It has been claimed that several degeneracies are present in a system which lead to strong
dephasing via two mechanisms. Within our approach the first mechanism \cite{Feldman} can be
interpreted as being based on the fact that there exists a set of quasi-degenerate states
which correspond to a shift of both edges.\cite{Wen}
One may parametrize them as following:
\begin{equation}
\frac{W_U\pi_U}{\sqrt{m}} = N+\frac{l}{m},\quad \frac{W_D\pi_D}{\sqrt{m}} = N-\frac{l}{m},
\label{shift}
\end{equation}
where $l = 0,\ldots, m-1$. We denote these states with $|l\rangle$,
and write $ |l+1\rangle =  e^{i(\phi_U+\phi_D)/\sqrt{m}}|l\rangle $.
The density matrix for the interferometer, with the QH edges in an equilibrium state,
can be written as
\begin{equation}
\rho_0 = \sum_{l=0}^{m-1} r_l|l\rangle\langle l|.
\label{rho0}
\end{equation}
The coherent part of the current generated by tunneling Hamiltonian (\ref{ham-gen})
and averaged with the density matrix (\ref{rho0}) reads
\begin{equation}
I_{LR} \propto \sum_l \sum_{\kappa,\kappa'}r_l
t_{L,\kappa}t^*_{R,\kappa'}\langle l| F^{\kappa-\kappa'}|l\rangle.
\label{shift-result}
\end{equation}
In our paper we use tunneling Hamiltonian (\ref{ham-0}) with $\kappa = \kappa'$. Therefore
the summation over the quantum number $l$ is trivial,  $\sum_lr_l =1$,
and does not lead to any physical effect. In contrast,
using additional Klein factors\cite{Feldman} implies that $\kappa-\kappa' = 1$.
Thus, in Eq.\ (\ref{shift-result}), the contribution from every shifted state
$|l\rangle$ acquires the phase factor $\langle l| F|l\rangle = e^{2\pi i l/m}$. Law {\it et al.} \cite{Feldman}
further assumed equal population $r_l=1/m$,
so that, after summation over $l$, the coherent part of the quasi-particle current vanishes.

In the argument sketched above, in addition to the specific choice
$\kappa-\kappa' = 1$ that has been addressed earlier, the assumption of the
degeneracy of states $|l\rangle$, or  equivalently, the high temperature limit $r_l = 1/m$,
is of crucial importance. We note, however, that the shift (\ref{shift}) leads to charging of the
edges, and the corresponding energy is not small. The experiments [\onlinecite{Heiblum2}-\onlinecite{mz7}]
have been done in the regime where the temperature and the bias are smaller than this
charging energy, and we do not see any difficulties, in principle, to achieve such a regime
in the fractional QHE case.

A second possible mechanism of dephasing is described in Ref.\ [\onlinecite{Stern}].
An additional Berry phase shift between two tunneling paths may appear if some
number $l$ of localized quasi-particles are present in the bulk. In this case
the coherent contribution to the tunneling current acquires the phase factor
$e^{2\pi i l/m}$, which is similar to the one encountered in the first mechanism of dephasing discussed above.
If the number, $l$, of quasi-particles fluctuates, then the coherent part of the quasi-particle
current may vanish,
$$
I_{LR}\propto\sum_{l} e^{2\pi i l/m}\to 0,
$$
as a result of averaging over these fluctuations.
We would like to stress that this mechanism may be avoided either by reducing the temperature,
so that the activation of spontaneous phase slips is slow, or by using a high-quality sample.
For instance, in the integer QHE case such telegraph processes have indeed been observed
in the experiment [\onlinecite{mz7}] and then reduced by tuning the system's parameters.

To conclude, we recall that the MZ interferometer, being a system that is strongly coupled to
Ohmic contacts, requires open boundary conditions. This leads to the non-commutativity
of tunneling Hamiltonians taken at different spatial points. We think that the non-commutativity
of spatially separated operators naturally follows from the topological character of the effective theory,
according to which the quasi-particles are not completely local objects, because they have ``tails''
in form of Wilson lines. Moreover, the non-locality in the effective theory does not contradict the
local character of the underlying microscopic theory which does not necessary
manifest itself in the low-energy limit.
The Klein factors, which have been introduced in earlier papers to
ensure the commutativity of tunneling Hamiltonians, require in fact closed boundary conditions and
are therefore not applicable to MZ interferometers.  Moreover, we have shown that
even when Klein factors may be used, they cannot guarantee the commutativity
of tunneling Hamiltonians in general, while the specific choice
proposed in Refs.\ [\onlinecite{Safi}] and [\onlinecite{Feldman}] needs further justification.
Finally, we remark that the strong suppression of the phase coherence suggested in Refs.\
[\onlinecite{Feldman}] and [\onlinecite{Stern}] is not a fundamental property of the fractional
QHE, and may be avoided in future experiments at sufficiently low temperatures.

\subsection{Periodicity in magnetic flux and modulation gate voltage}
\label{flux-period}

\begin{figure}[hbt]
\epsfxsize=8cm
\begin{center}
\epsfbox{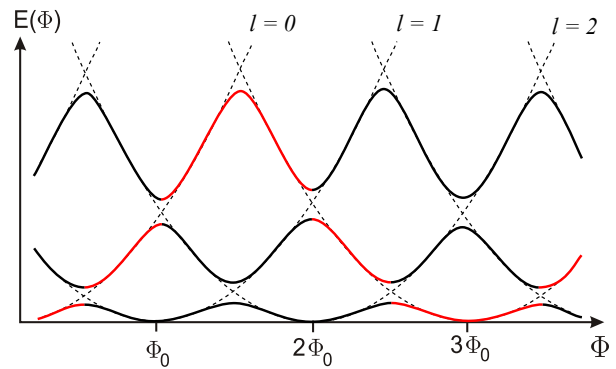}
\end{center}
\caption{(Color online) The energy spectrum of a QH fluid at $\nu=1/3$ in a Corbino disk
is schematically shown. The dashed lines show the Coulomb charging energy
of an isolated QH fluid, as a function of the magnetic flux $\Phi$ threading
the Corbino disk. Different
branches correspond to different numbers of quasi-particles at the edges.
In the presence of inter-edge quasi-particle tunneling, and for weak coupling
to metallic reservoirs, an energy gap opens at the degeneracy points, where
different branches intersect. If the flux varies adiabatically, the QH fluid
follows the ground state, so that the energy is a periodic function of the flux
with the electronic period $\Phi_0$. If the variation of the flux is relatively
fast (quasi-particle tunneling is weak), the QH fluid follows the ``electronic''
branch (shown in red) along the lines l=0 and l=3, so that the number of electrons at the edge changes by one.
In this case the energy oscillates with the quasi-particle period $3\Phi_0$.}
\vspace{-2mm}
\label{magn}
\end{figure}

The purpose of this discussion is to clarify the origin of the quasi-particle periodicity
in physical observables. Let us first analyze the dependence of the energies of the zero
modes eigenstates $|l\rangle$ (see Appendix \ref{Klein}) of a QH fluid at $\nu=1/m$
on the singular magnetic flux $\Phi$ threading the Corbino disk. This problem has
been considered by Thouless and Gefen in Ref.\ [\onlinecite{Thouless}]. Following
their argument, we note that the main contribution to the energy comes
from Coulomb interactions. Therefore, under variation of the flux $\Phi$,
the energies of zero modes eigenstates of a QH fluid, isolated inside a Corbino disk, follow
the branches with a fixed number of electrons. However, if the Corbino disk is weakly
coupled to metallic reservoirs, the number of electrons
in the QH fluid is not conserved. Therefore, if the ground state energy initially grows
with the flux $\Phi$, it then switches to another branch by changing the number of electrons
by 1 and starts to decrease with the flux. This behavior (shown by the red line in
Fig.\ \ref{magn}) repeats periodically with a period equal to $m\Phi_0$.

The electronic periodicity is restored if one takes into account the possibility
of quasi-particle tunneling between the inner and outer edges of the Corbino disk.
Such a perturbation mixes the states $|l\rangle$, with $l=0,\ldots,m-1$ and opens
a gap at the degeneracy points (see Fig.\ \ref{magn}). As a result, under an adiabatic
variation of the magnetic flux, the QH fluid will follow the lowest energy state
with the electronic period $\Phi_0$.

Far away from equilibrium, when a potential difference is applied to metallic reservoirs
and a charge current flows through an MZ interferometer, the physical situation
is generally more complicated. However, one  can still rely on the effective theory by taking
into account the described above dependence of the energy of zero modes on the magnetic flux
and writing the Hamiltonian as
\begin{equation}
\mathcal{H} = \sum_{s = U,D}\frac{v_s}{4\pi}\int dx
\left(\partial_x\phi_s + \frac{2\pi}{W_s\sqrt{m}}\frac{\Phi}{\Phi_0}\right)^2.
\label{ham-shift}
\end{equation}

We stress that the the insertion of the singular flux does not affect the AB phase.
This follows form the fact that there are two contributions to the integral
$\delta\varphi_{AB}=(1/\sqrt{m})\int_\gamma dr^\mu\delta b_\mu $ over the MZ contour
$\gamma$. One contribution arises from the change in the field $b_\mu$ due to the variation of the
external field $A_\mu$, as follows from relation (\ref{BA}).
The second contribution comes from the variation of the total charge inside the MZ contour,
repelled by the singular flux, according to relation (\ref{j-bulk}).
These two effects cancel each other exactly, which can be viewed as a result of
screening of the external field $A_\mu$ by the QH system.

To determine the current through the interferometer the averaging in the linear-response
formula (\ref{interf}) must be done with respect to the equilibrium density matrix
$\rho_0 = e^{-\beta\mathcal{H}}/Z_0$.
In the thermodynamic limit, the fields $\partial_x\phi_s$
can be shifted when performing the averaging so as
to eliminate the dependence on the magnetic flux from Hamiltonian (\ref{ham-shift}).
This implies that the only physical consequence of the insertion of the singular magnetic flux
in a finite size Corbino disk is the Coulomb blockade effect which has an electronic periodicity.
However, this effect
is absent in MZ interferometers due to the strong coupling to Ohmic contacts.

The  modulation of the AB phase via the insertion of a point-like flux is an idealization.
In a typical experiment, the magnetic field can be changed only  uniformly. In this case the phase
acquired by a quasi-particle is no longer a topological number and may depend, e.g.,
on the processes in Ohmic contacts in a complicated way. Therefore, it appears to be more appropriate
to investigate the periodicity of AB oscillations by changing the modulation gate voltage instead
of changing the magnetic field.

A gate voltage applied at the edge of a QH liquid leads to the displacement $y(x)$
of the edge at the point $x$. This displacement may be described as an accumulation
of background 1D charge densities, $\delta\rho_i(x) = \sigma_iQ_i^2y(x)/2\pi l_B^2$, in the edge
channels. In the presence of these densities, the fields $\phi_i$ are redefined as
$\partial\phi_i \to \partial\phi_i + \sigma_iQ_i y(x)/l_B^2$. As a consequence, the correlation function
$\langle\psi^\dag(x,t)\psi(0,0)\rangle$ of an excitation $\psi = e^{ip_i\phi_i}$ acquires a phase shift
$\delta\varphi_0 = \sigma_ip_iQ_i\int_0^x dx' y(x') /l_B^2$. This phase shift is proportional to
the charge of the quasi-particle, $Q_{\rm em}(\mathbf{n}) = \sum_i\sigma_ip_iQ_i$, and to the
area of the deformation $S = \int_0^x dx' y(x')$. Therefore the phase shift may be interpreted
as an AB phase:
\begin{equation}
\delta\varphi_0(\mathbf{n}) = Q_{\rm em}(\mathbf{n})\cdot S/l_B^2
= 2\pi Q_{\rm em}(\mathbf{n})\frac{\delta\Phi}{\Phi_0},
\label{add-phas}
\end{equation}
where $\delta\Phi$ is the variation of the magnetic flux through the closed path of the MZ interferometer,
resulting from the deformation. For simplicity of notations, this additional phase (\ref{add-phas})
is included in the tunneling amplitudes in Eq.\ (\ref{phas-ham}).

To conclude, in our model, the periodicity of AB oscillations in the average current is determined \
by the charges of quasi-particles that tunnel. To verify this prediction, one should compare the periodicity
in the weak tunneling regime, where only electron tunneling is possible, with the one in the weak backscattering
regime, where the tunneling of quasi-particles is most relevant.

\section{Chern-Simons theory and illustration of holography}
\label{cs-bulk}

In this appendix we illustrate the holographic principle by constructing the bulk effective models
and showing that they determine the minimal edge models discussed in Sec.\ \ref{sec:multi-field-edge}.
We assume that the edge currents originate as deformations of incompressible fluids. These fluids
are described by a family of separately conserved bulk currents, $j_{i\mu}$, with $\partial_\mu j_i^\mu = 0$.
We solve the continuity equations by introducing potentials $b_{i\mu}$,
\begin{equation}
j_{i\mu} = \frac{1}{2\pi}\epsilon_{\mu\nu\lambda}\partial^\nu b_i^\lambda,
\label{j-bulk}
\end{equation}
where the Einstein summation convention is assumed. The currents are invariant under the gauge transformations
$b_{i\mu}\to b_{i\mu} +\partial_\mu f_i$. By counting dimensions, it is easy to see that the gauge
invariant action for these potentials,
\begin{equation}
S_{\rm bulk}[b_i] = (1/4\pi)\sum_i\sigma_i\int_{D}d^3r \epsilon_{\mu\nu\lambda}b^\mu_i\partial^\nu b^\lambda_i,
\label{bulk-act}
\end{equation}
has zero dimension, while all other possible terms have {\em lower} dimensions, i.e., are {\em irrelevant} at low energies.
For example, the Maxwell-like action has dimension $-1$.

The total electric current can be written as a linear combination of incompressible currents
$j_{\rm em}^{\mu} = \sum_iQ_ij_i^\mu$. Hence the term in the action describing the interaction
with an external electromagnetic field is:
\begin{multline}
S_{\rm int}[b_i,A] = \int_{D}d^3r A_\mu j^\mu_{\rm em} \\
= (1/2\pi)\sum_i\int_{D}d^3r A_\mu Q_i\epsilon^{\mu\nu\lambda}\partial_\nu b_{i\lambda}.
\end{multline}
Integrating out the fields $b_{i\mu}$, we arrive at an effective action for the electromagnetic
field in the Chern-Simons form:
\begin{equation}
S_{\rm eff}[A] = (1/4\pi)\sum_i\sigma_iQ_i^2 \int_D d^3r  \epsilon^{\mu\nu\lambda}A_\mu\partial_\nu A_\lambda.
\end{equation}
Comparing this result to Eq.\~(\ref{canc-2}), we conclude that the constraint on the coupling
constants $Q_i$ is the same
as in the edge theory, namely $\sum_i\sigma_iQ_i^2 = \nu$.

Action (\ref{bulk-act}) appears in the context of topological theory, where excitations are given
by Wilson lines. \cite{CS-theory}
For instance, a general local excitation at the point $r$ may be written as:
\begin{equation}
\psi_q(r) = \exp\Big(i \sum\limits_jq_j\!\int\limits^r_{r_0}dr^\mu b_{j\mu}\Big).
\label{op-bulk}
\end{equation}
The statistical phase of two excitations of type (\ref{op-bulk}) is given by braiding of
the corresponding Wilson lines. \cite{CS-theory} Considering two excitations labeled with $q_{1j}$
and $q_{2j}$ we arrive, after a simple calculation of braiding, at the following expression for
the statistical phase: $\theta_{12} = \pi\sum_i\sigma_j q_{1j}q_{2j}$. It is important that this
expression coincides with the one for the edge excitations. Moreover,
if we define the charge operator as an integral over a space-like plane
$Q_{\rm em} = (Q_i/2\pi)\int d^2r  \epsilon_{\nu\lambda}\partial^\nu b_i^\lambda $,
then the charge of the excitation (\ref{op-bulk}) is
$Q_{\rm em} = \sum_i\sigma_iQ_iq_i$, i.e.,\ it takes the same form as in the edge theory.

The coincidence of bulk and edge expressions for charges and statistical phases of excitations
illustrates the holographic principle at work in QH systems.
Indeed, it is easy to see that the whole classification of effective models at the edge
applies also to the bulk. Furthermore, Fr\"{o}hlich and Pedrini \cite{exact} proposed an exact mapping
between edge and bulk models, assuming that the edge excitations originate from incompressible
deformations of QH liquids. Below we summarize the main steps of this construction.

\begin{figure}[hbt]
\epsfxsize=7cm
\begin{center}
\epsfbox{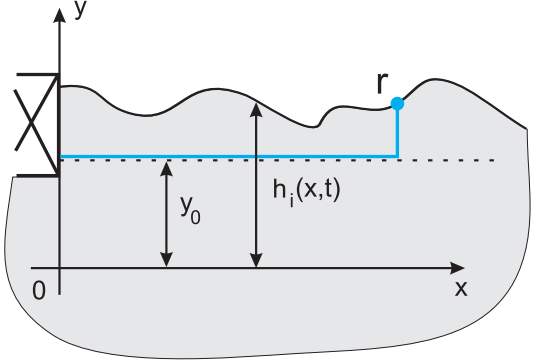}
\end{center}
\caption{(Color online) The construction of edge excitations as incompressible deformations of a QH liquid.
The Wilson line (drawn in blue) starts from the Ohmic contact (drawn in white) and
goes to the physical edge $y=h_i(x,t)$ of the $i{\rm th}$ QH liquid (solid line) along
the auxiliary boundary at $y = y_0$ (dashed line), and ends at the point r.} \vspace{-2mm}
\label{e-b}
\end{figure}

The key idea is to represent the insertion of an operator $\exp\big(i\sum_j\!q_j\phi_j\big)$
at the edge as insertion of a Wilson line (\ref{op-bulk}) of the bulk theory
by placing the end point $r$ at the boundary and $r_0$ in the Ohmic contact. To realize this idea
we choose a particular Wilson line shown in Fig.\ \ref{e-b}.
Parametrizing incompressible deformations of QH liquids by the set of
functions $y = h_i(x,t)$, one may introduce the boundary fields:
\begin{equation}
\phi_i(x,t) = \int\limits^{h_i(x,t)}_{y_0}dy' b_{iy}(x,y',t) + \int\limits_{x_0}^x dx' b_{ix}(x',y_0,t),
\label{edge-f}
\end{equation}
where $y=y_0$ is the auxiliary boundary and $x_0$ is the coordinate of the Ohmic contact.

Integrating out the bulk fields $b_{i\mu}$, we find that
\cite{footnote-x1}
\begin{equation}
b_{i\mu} = \sigma_iQ_iA_\mu.
\label{BA}
\end{equation}
Taking the derivative of Eq.~(\ref{edge-f}) with respect to $x$, one obtains
$\sigma_i\partial_x\phi_i = Q_i A_y(x,h_i,t) \partial_xh_i +Q_i\int_{y_0}^{h_i}dy'\partial_xA_y(x,y',t) + Q_iA_x(x,y_0,t)$.
In the low-energy limit we may neglect the term proportional to $\partial_xh_i$.
Then, taking into account that the magnetic field is constant $\partial_xA_y - \partial_yA_x = 1/l_B^2$,
we relate the derivatives $\partial_x\phi_i$ to the edge densities\cite{footnote-x2}
$\rho_i(x,t) = \sigma_iQ_i^2[h_i(x,t)-y_0]/2\pi l_B^2$:
\begin{equation}
\rho_i(x,t) = \frac{1}{2\pi}\big[Q_i\partial_x\phi_i + \sigma_iQ_i^2A_x(x,h_i,t)\big].
\end{equation}
This expression for the charge densities of the edge channels coincides with the one derived from
action (\ref{act-2}), provided we further assume that $h_i$ is small in the low-energy limit
and replace $A_x(x,h_i,t)\to A_x(x,y_0,t)$. In other words, we assume linear coupling
to the electromagnetic field.

The continuity equations $\dot{\rho_i}+\partial_x\int_{y_0}^{h_i}j_{ix} = j_{iy}$,
after the substitution of Eq.\ (\ref{j-bulk}),
take the form:
\begin{multline}
\frac{Q_i\dot{h_i}}{2\pi l^2} + \partial_x\int_{y_0}^{h_i}dy'[\partial_{y'} b_{it}(x,y',t)-\partial_tb_{iy}(x,y',t)]
\\= \partial_x b_{it}(x,y_0,t)-\partial_tb_{ix}(x,y_0,t).
\label{cont}
\end{multline}
Linearizing this equation with respect to $h_i$, we find the equations of motion
for the boundary fields $\sigma_i\partial_t D_x\phi_i = - Q_i \partial_xa_t$,
where $a_\mu$ is the boundary value of the electromagnetic vector potential, and
$D_\mu\phi_i = \partial_\mu\phi_i + \sigma_iQ_ia_\mu$ are covariant derivatives.
These equations of motion can be derived from the action
\begin{equation}
S \propto \sum_i\int dtdx (\sigma_iD_t\phi_iD_x\phi_i + Q_i\epsilon_{\mu\nu}a^\mu\partial^\nu\phi_i),
\end{equation}
which agrees with action (\ref{act-2}) in the case $v_i=0$.
The velocities of edge modes in the above analysis are zero,
because we have not taken into account the confining potential and the
interaction effects at the edge.

It is important to note that the bulk-edge correspondence described above does not always hold.
For instance, compressible strips may be present at the edge. \cite{strips} In this case, the edge theory
will contain more fields than the bulk theory. Nevertheless, bulk excitations should be always present
in the edge spectrum. In the present work we only consider minimal models of the edge, for the following reasons. First of
all, in the case when the number of chiral fields at the edge is large, scaling dimensions of extra electrons
associated with these fields are usually large, \cite{Fr-abel} i.e., these excitations are not observable
in the low-energy limit. Second, non-chiral models with additional fields are likely to be unstable against
disorder. \cite{PKF} However, the possible relevance of models with a large number of channels deserves a separate,
detailed analysis.

\section{Mathematical aspects of the theory of QH lattices}
\label{sec:frohlich-theory}

In this section we wish to briefly describe the QH lattice construction proposed in
Refs.\ [\onlinecite{Fr-abel,exact}].
This construction is a mathematical reformulation of physical requirements for the effective
theory of a QH system discussed in Secs.\ \ref{bulk-edge} and \ref{sec:multi-field-edge}.
It provides a method, based on using invariants of lattices, to classify physically allowed
low-energy effective models of QH edge states. The main physical consequences of the lattice
construction include a determination of the minimal charge of quasi-particle, of the minimal number of
edge channels, for given filling factor, etc. Here we summarize the lattice construction and
present the results without proof.

First, we recall that action (\ref{act-2}) of the effective theory is parameterized by the
vector of coupling constants, $Q_i$, introduced in Sec.\ \ref{sec:multi-field-edge}. Local
excitations are represented by vertex operators, $\exp\big(i\sum_jq_j\phi_j\big)$, and labeled by
vectors $\mathbf{q} = \{q_i\}$; so the sum of two such vectors corresponds to the product of operators. Defining the scalar product:
\begin{equation}
\langle\mathbf{a},\mathbf{b}\rangle = \sum_ia_ib_i\sigma_i,
\end{equation}
where $\sigma_i$ is the chirality of $i{\rm th}$ channel, we represent the electric charge (\ref{ch-def})
of an excitation corresponding to the vector $\mathbf{q}$ and the statistical phase (\ref{ph-def}) of two excitations corresponding to $\mathbf{q}_1$ and $\mathbf{q}_2$ as
\begin{equation}
Q_{\rm em} = \langle\mathbf{Q},\mathbf{q}\rangle, \quad \theta_{12} = \pi \langle\mathbf{q}_1,\mathbf{q}_2\rangle,
\end{equation}
respectively.

As discussed in Sec.\ \ref{sec:multi-field-edge}, after fixing the coupling constants $Q_i$
(i.e.\ fixing the action), we have to choose electronic excitations. We denote them with $\mathbf{q}_\alpha$. Multi-electron excitations form an integral lattice $\Gamma$:
\begin{equation}
\Gamma = \{k_\alpha\mathbf{q}_\alpha | k_\alpha\in \mathbb{Z}\}.
\end{equation}
It has been mentioned in Sec.\ \ref{sec:excitations-theory} that choosing some different sets of
elementary electronic excitations is equivalent to choosing different bases in the same lattice $\Gamma$. Thus, the lattice $\Gamma$ describes an effective theory in a
basis-independent way. The condition that the electric charge $\langle\mathbf{Q},\mathbf{q}\rangle$
of any combination of electrons $\mathbf{q}\in\Gamma$ is integer implies that the
vector of couplings $\mathbf{Q}$ belongs to the dual lattice $\Gamma^*$. Thus, by choosing
the lattice $\Gamma$ and a vector $\mathbf{Q}$ of its dual, one selects a
particular effective model.

Next, important physical constraints on the effective models discussed in Sec.\ \ref{bulk-edge}
can be formulated as following:
\begin{itemize}
\item The condition of anomaly cancellation, for a given filling fraction $\nu$, implies that
$\langle\mathbf{Q},\mathbf{Q}\rangle = \nu$.
\item The correct charge of electronic operators is guaranteed if the greatest common divisor
of the coordinates of $\mathbf{Q}$ in $\Gamma^*$ is equal to $1$, because this divisor is
equal to the minimal value of $\langle\mathbf{Q},\mathbf{q}\rangle$ for $\mathbf{q}\in\Gamma$.
\item The correct statistical phase of electronic excitations is a consequence of the
condition $\langle\mathbf{Q},\mathbf{q}\rangle \equiv \langle\mathbf{q},
\mathbf{q}\rangle\hspace{5pt} ({\rm mod}\hspace{2pt} 2)$, $\forall \mathbf{q}\in\Gamma$.
\end{itemize}

The spectrum of allowed local excitations follows from the requirement that the wave function of the QH state is single-valued in the presence of an excitation
$\exp\big(i\sum_j\!p_j\phi_j\big)$. We mentioned in Sec.\ \ref{bulk-edge} that this condition is
equivalent to having integer relative statistical phases between electrons and quasi-particles:
\begin{equation}
\langle\mathbf{p},\mathbf{q}_\alpha\rangle\in \mathbb{Z}.
\end{equation}
Thus, the lattice of allowed excitations is $\Gamma^*\supseteq\Gamma$.
As we have shown in Sec.\ \ref{sec:multi-field-edge}, the scaling dimension, $\Delta$, of the correlation
function of the excitation $\mathbf{p}$ does not depend on the Hamiltonian in a purely chiral theory.
It is equal to the statistical phase $\Delta = \langle\mathbf{p},\mathbf{p}\rangle$. So, for
purely chiral models, the pair $(\Gamma,\mathbf{Q})$ provides complete information about the effective
theory. In order to do explicit calculations, one needs to introduce a particular basis for $\Gamma$.
It is, however, not trivial to verify whether two different bases generate the same lattice. To
distinguish effective models and, therefore, classify them, one needs basis-independent
information about lattices. Such information is provided by {\em lattice invariants}.

To classify the pairs $(\Gamma,\mathbf{Q})$ satisfying the conditions discussed above,
we introduce most important lattice invariants. Obvious invariants are the filling factor
$\nu = n_H/d_H$, with $n_H$ and $d_H$ as coprime integers, and the dimension (or rank) of the lattice
$N = \dim \Gamma$. One may show that, for any basis, $\{\mathbf{e}_\alpha\}$, of $\Gamma$, the determinant
of the Gram matrix $\Delta_\Gamma = \det \langle \mathbf{e}_\alpha,\mathbf{e}_\beta\rangle$ is also
an invariant. \cite{fr-inv} An interesting property of this determinant is the factorization
$\Delta_\Gamma = l d_H$, where $l$ is an integer number called {\em level}. Moreover, one may show
that, for any basis $\{\mathbf{e}^\alpha\}$ of $\Gamma^*$, the greatest common divisor $g = \gcd(Q^1,\ldots,Q^N)$
of the numbers  $Q^\alpha = \Delta_\Gamma\langle\mathbf{Q},\mathbf{e^\alpha}\rangle$ is an invariant, and that
$l = \lambda g$, where $\lambda$ is an integer. This number $\lambda$ is an important
parameter often called the {\em charge parameter}. It determines the minimal possible electric charge of a quasi-particle:
\begin{equation}
e^* = \min_{\mathbf{p}\in\Gamma^*,\langle\mathbf{Q},\mathbf{p}\rangle \neq 0}|\langle\mathbf{Q},
\mathbf{p}\rangle| = \frac{1}{\lambda d_H}.
\end{equation}

Finally, one introduces two further invariants, called minimal and maximal relative angular momenta. The minimal relative angular momentum is defined as
\begin{equation}
\ell_{\min} = \min_{\mathbf{q}\in\Gamma,\langle\mathbf{Q},\mathbf{q}\rangle =1}\langle\mathbf{q},\mathbf{q}\rangle.
\end{equation}
Introducing the set, $B_{\mathbf{Q}}$, of all possible electronic bases, $\{\mathbf{q}_\alpha\}$, of
$\Gamma$, for a given $\mathbf{Q}$ (i.e., such that $\forall\alpha: \langle\mathbf{Q},\mathbf{q}_\alpha\rangle = 1$),
one defines the maximal relative angular momentum as
\begin{equation}
\ell_{\max} = \min_{\{\mathbf{q}_\alpha\}\in B_{\mathbf{Q}}}
\left(\max_\alpha\langle\mathbf{q}_\alpha,\mathbf{q}_\alpha\rangle\right).
\end{equation}
The invariants $\ell_{\min}$ and $\ell_{\max}$ cannot take arbitrary values. For instance,
in purely chiral models, they are constrained by the inequality $1/\nu\leq\ell_{\min}\leq\ell_{\max}$.

The simplest examples of such lattice construction appear in the case of dimension $N = 1$.
For one-dimensional lattices, there is only one independent invariant
$\ell_{\max} = \ell_{\min} = \Delta_\Gamma = m$, where $m$ is an odd integer.
This number is nothing but the statistical phase of an electron, therefore
$Q = 1/\sqrt{m}$ and the filling factor $\nu = 1/m$. Moreover, it is easy to see
that the level $l = \Delta_\Gamma/d_H = 1$, hence the charge parameter $\lambda=1$,
and we find that the minimal electric charge is $e^*= 1/m$. We conclude that the QH lattices
\begin{equation}
\Gamma_m = \{n\sqrt{m}\mathbf{e}| n\in\mathbb{Z}\}
\end{equation}
with coupling $\mathbf{Q}_m = \mathbf{e}/\sqrt{m}$ and $\nu = 1/m$ are the only ones
allowed for $N=1$.

For the two-field case, $N=2$, the construction of allowed lattices is more complex. From the definition of relative angular momenta
it follows that we may choose electronic bases with a Gram matrix of the following form:
\begin{equation}
K = \left(
      \begin{array}{cc}
        \ell_{\min} & b \\
        b & \ell_{\max} \\
      \end{array}
    \right).
    \label{gr-1}
\end{equation}
We consider lattices with $\ell_{\max} < 7$, which are physically most relevant. \cite{footnote7}
Then one may simply list all models by going through all possible values of $\ell_{\max}$,
$\ell_{\min}$ and $b$. \cite{Fr-abel} Furthermore, we limit our attention to the case $\ell_{\min} = \ell_{\max}$,
which is most important in the context of our paper.

For convenience, we choose coordinates such that $\mathbf{Q} = (\sqrt{\nu},0)$.
The condition of unit charge, $\langle\mathbf{Q},\mathbf{q}_\alpha\rangle = 1$,
for $\alpha =1,2$, partially fixes the form of electron vectors, $\mathbf{q}_\alpha$, in these coordinates.
Namely, $\mathbf{q}_1 = (1/\sqrt{\nu},s)$ and $\mathbf{q}_2 = (1/\sqrt{\nu},-s)$,
where the number $s$ is yet to be determined. It follows from the requirement
$|\mathbf{q}_1|^2 = |\mathbf{q}_2|^2 = \ell_{\max}$ that $s^2 = \ell_{\max}-1/\nu$.
The mutual statistical phase of two electrons, $\theta_{12} = \pi (1/\nu-s^2)$, should be an integer.
This implies that $\ell_{\max}+\theta_{12}/\pi = 2/\nu$ is an integer number.
Thus we see that the special case $\ell_{\min} = \ell_{\max}$ corresponds to
$\nu = 2/m$ where $m$ is an integer. Note that, for $\ell_{\max}<7$, the converse statement
is also true, i.e., for $\nu = 2/m$, {\em all} the two dimensional lattices have $\ell_{\min} = \ell_{\max}$.

After some elementary calculations, we find that, for $\nu = 2/m$, Gram matrix (\ref{gr-1}) may be expressed as:
\begin{equation}
K = \left(
      \begin{array}{cc}
        \ell_{\max} & \ell_{\max}-l \\
        \ell_{\max}-l & \ell_{\max} \\
      \end{array}
    \right).
    \label{gr-2}
\end{equation}
From condition (\ref{cond-main}), applied to Eq.~(\ref{gr-2}), we find that the level $l = 2\ell_{\max}-m = 2s^2$.
Therefore, all the lattices in the case $\nu = 2/m$ can be parameterized by {\em only two} numbers, $m$ and $l$.
It is important to observe that, for all these lattices, $\lambda = 1$. Hence, for two-field models with $\nu = 2/m$,
the minimal charge is always $e^* = 1/m$. Introducing a pair of orthogonal vectors, $\mathbf{e}_{1,2}$,
so that $\mathbf{q}_{1,2} = \sqrt{m/2}\mathbf{e}_1\pm\sqrt{l/2}\mathbf{e}_2$,
the lattices may be written explicitly as
\begin{multline}
\Gamma_{ml} = \{\sqrt{m/2}(n_1+n_2)\mathbf{e}_1 \\
+ \sqrt{l/2}(n_1-n_2)\mathbf{e}_2| n_{1,2}\in\mathbb{Z}\},
\end{multline}
and the couplings $\mathbf{Q}_{ml} = \sqrt{2/m}\,\mathbf{e}_1$. Finally, we stress that $K$-matrices (\ref{eq:25})
and (\ref{eq:33}) proposed in Sec.\ \ref{sec:simplest-two-field} are exactly the Gram matrices (\ref{gr-2}) for the
particular case $\nu = 2/3$.


\begin{thebibliography}{99}

\bibitem{QHE} {\em The Quantum Hall Effect}, edited by R.E. Prange and S.M.
Girvin (Springer, New York, 1987).

\bibitem{Datta}
S. Datta, {\em Electronic Transport in Mesoscopic Systems} (Cambridge University Press, Cambridge, 1999).

\bibitem{hologr}
G.~'t Hooft, Proceedings of the Salamfest, 0284-0296 (1993);
L.~Susskind, J.\ Math.\ Phys.\ {\bf 36}, 6377 (1995).

\bibitem{Witten}
The well known examples of the holographic principle in mathematical physics are described in the papers:
E. Witten, Adv. Theor. Math. Phys. {\bf 2}, 253 (1998);
O.~Aharony, S.~S.~Gubser, J.~M.~Maldacena, H.~Ooguri and Y.~Oz,
Phys.\ Rept.\ {\bf 323}, 183 (2000).

\bibitem{inflow}
Examples of the anomaly inflow at the edge in some other physical systems are provided by
J.~A.~Harvey and O.~Ruchayskiy, J. High Energy Phys. 2001, 044;
A.~Boyarsky, O.~Ruchayskiy and M.~Shaposhnikov, Phys.\ Rev.\ D {\bf 72} 085011 (2005).


\bibitem{Wen}
X.-G. Wen, Phys. Rev. B {\bf 41}, 12838 (1990).

\bibitem{Frol}
J. Fr\"{o}hlich, A. Zee, Nucl. Phys. B{\bf 364}, 517 (1991); J. Fr\"{o}hlich and T. Kerler,
Nucl. Phys. B{\bf 354}, 369 (1991).

\bibitem{Fr-abel}
J. Fr\"{o}hlich, U.M. Studer, E. Thiran, J. Stat. Phys. {\bf 86}, 821 (1997).

\bibitem{Fr-non-abel}
J. Fr\"{o}hlich, B. Pedrini, C. Schweigert and J. Walcher, J. Stat. Phys. {\bf 103}, 527 (2001).

\bibitem{wen-iv}
X.-G. Wen, Phys. Rev. B {\bf 44}, 5708 (1991).

\bibitem{Chang-West}
A.M. Chang, L.N. Pfeiffer, and K.W. West,
Phys. Rev. Lett. {\bf 77}, 2538 (1996).

\bibitem{Gr-Chang}
M. Crayson, D.C. Tsui, L.N. Pfeiffer, K.W. West, and A.M. Chang, Phys. Rev. Lett. {\bf 80}, 1062 (1998).

\bibitem{podborka}
For a review, see
A.M. Chang, Rev. Mod. Phys. {\bf 75}, 1449 (2003).

\bibitem{strips}
D.B. Chklovskii, B.I. Shklovskii, and L.I. Glazman,
Phys. Rev. B {\bf 46}, 4026 (1992).

\bibitem{PKF}
C.L. Kane, M.P.A. Fisher and J. Polchinski, Phys. Rev. Lett. {\bf 72}, 4129 (1994); C.L. Kane and
Matthew P.A. Fisher, Phys. Rev. B {\bf 51}, 13449 (1995).

\bibitem{phonon}
S. Khlebnikov, Phys. Rev. B {\bf 73}, 045331 (2006).

\bibitem{noise-prop}
C.L. Kane and Matthew P.A. Fisher, Phys. Rev. Lett. {\bf 72}, 724 (1994)

\bibitem{fract-charge-meas}
L. Saminadayar, D.C. Glattli, Y. Jin and B. Etienne, Phys. Rev. Lett. {\bf 79}, 2526 (1997);
R. de-Picciotto, Nature, {\bf 389}, 162 (1997).

\bibitem{quarter-charge}
M. Dolev, M. Heiblum, V. Umansky, A. Stern, and D. Mahalu, Nature {\bf 452}, 829 (2008).

\bibitem{Goldman}
F.E. Camino, W. Zhou and V.J. Goldman, Phys. Rev. Lett. {\bf 95}, 246802 (2005).

\bibitem{mz1}
Y.\ Ji, Y. Chung, D. Sprinzak, M. Heiblum, D. Mahalu, and H.
Shtrikman, Nature (London) {\bf 422}, 415 (2003).

\bibitem{Safi}
R. Guyon, P. Devillard, T. Martin and I. Safi, Phys. Rev. B {\bf 65}, 153304 (2002).

\bibitem{Kane}
C.L. Kane, Phys. Rev. Lett. {\bf 90}, 226802 (2003).

\bibitem{Feldman}
K.T. Law, D. E. Feldman and Y. Gefen, Phys. Rev. B {\bf 74}, 045319 (2006).

\bibitem{Vishvesh}
E.-A. Kim, M.J. Lawler, S. Vishveshwara and E. Fradkin, Phys. Rev. B {\bf 74}, 155324 (2006).

\bibitem{Averin}
V.V. Ponomarenko and D.V. Averin, Phys. Rev. Lett. {\bf 99}, 066803 (2007).

\bibitem{Heiblum2}
I. Neder, M. Heiblum, Y. Levinson, D. Mahalu, and V. Umansky, Phys. Rev. Lett. {\bf 96}, 016804 (2006);
I. Neder, F. Marquardt, M. Heiblum, D. Mahalu, and V. Umansky, Nat. Phys. {\bf 3}, 534 (2007).

\bibitem{Basel}
E. Bieri, Ph.D. thesis, University of Basel, 2007; E. Bieri, M. Weiss, O. Goktas,
M. Hauser, S. Oberholzer, and C. Schonenberger, Phys. Rev. B {\bf 79}, 245324 (2009).

\bibitem{Glattli1}
P. Roulleau, F. Portier, D.C. Glattli, P. Roche, A. Cavanna,
G. Faini, U. Gennser, and D. Mailly, Phys. Rev. B {\bf 76}, 161309(R) (2007); Phys. Rev. Lett. {\bf 100}, 126802 (2008).

\bibitem{mz7}
L.V. Litvin, H.-P. Tranitz, W. Wegscheider, and C. Strunk, Phys. Rev. B {\bf 75}, 033315 (2007);
L.V. Litvin, A. Helzel, H.-P. Tranitz,
W. Wegscheider, and C. Strunk, Phys. Rev. B {\bf 78}, 075303 (2008).

\bibitem{Sukh-Che}
E.V. Sukhorukov and V.V. Cheianov, Phys. Rev. Lett. {\bf 99},
156801 (2007).

\bibitem{Chalker}
J.T. Chalker, Y. Gefen, and M.Y. Veillette,
Phys. Rev. B {\bf 76}, 085320 (2007).

\bibitem{Neder}
I. Neder and E. Ginossar, Phys. Rev. Lett. {\bf 100}, 196806 (2008).

\bibitem{Sim}
S.-C. Youn, H.-W. Lee, and H.-S. Sim, Phys. Rev. Lett. {\bf 100}, 196807 (2008).

\bibitem{our}
I.P. Levkivskyi and E.V. Sukhorukov, Phys. Rev. B {\bf 78}, 045322 (2008).

\bibitem{footnote2}
We define the statistical phase $\theta_{12}$ of two operators $\psi_1$ and $\psi_2$
via the relation $\psi_1(x)\psi_2(x') = e^{i\theta_{12}}\psi_2(x')\psi_1(x)$. The
microscopic construction \cite{micro} shows that so defined statistical phase indeed
takes integer values for single valued excitations of the Laughlin state. \cite{Laugh}

\bibitem{Jackiw}
S.B. Treiman, R. Jackiw, D.J. Gross, {\it Lectures on Current Algebra and its Applications},
(Princeton University Press, Princeton N.J., 1972).

\bibitem{wen-text}
X.-G. Wen,
{\em Quantum Field Theory of Many-body Systems},
(Oxford University Press, New York, 2007).

\bibitem{boson-textbook}
A.O. Gogolin, A.A. Nersesyan and A.M. Tsvelik, {\em Bosonization and Strongly Correlated Systems}
(Cambridge University Press, Cambridge, 1998).

\bibitem{Laugh}
R.B. Laughlin, Phys. Rev. Lett. {\bf 50}, 1395 (1983).

\bibitem{footnote-ckm}
In fact, the introduced matrix $q_{\alpha i}$ is an analogue of the well known
Cabibbo-Kobayashi-Maskawa matrix in particle physics.

\bibitem{footnote3}
Note that introducing new fields $\chi_\alpha = \sum_iq_{\alpha i}\phi_i$, one arrives at the action
$S = \int dtdx [K^{-1}_{\alpha\beta}\partial_t\chi_\alpha\partial_x\chi_\beta
+ V_{\alpha\beta}\partial_x\chi_\alpha\partial_x\chi_\beta]$,
which is often used in the literature. However, this form of the action is not
convenient for the classification of effective models and may lead to a number of confusions.

\bibitem{footnote7}
Numerical simulations \cite{number-w} show that for larger values of statistical phases the QH state
is not stable,
and electrons form a Wigner crystal. \cite{QHE}

\bibitem{wen23}
X.G. Wen, Phys. Rev. Lett. {\bf 64}, 2206 (1990);
A.H. MacDonald, Phys. Rev. Lett. {\bf 64}, 220 (1990);
M.D. Johnson, A.H. MacDonald, Phys. Rev. Lett. {\bf 67}, 2060 (1991).

\bibitem{jain}
J.K. Jain, Phys. Rev. Lett. {\bf 63}, 199 (1989).

\bibitem{num}
Z.-X. Hu, H. Chen, K. Yang, E. H. Rezayi, X. Wan, Phys. Rev. B {\bf 78}, 235315 (2008).

\bibitem{tilted}
R.G. Clark, S.R. Haynes, J.V. Branch, A.M. Suckling, P.A. Wright, P.M.W. Oswald, J.J. Harris, and C.T. Foxon,
Surf. Sci. {\bf 229}, 25 (1990); J.P. Eisenstein, H.L. Stormer, L.N. Pfeiffer, and K.W. West,
Phys. Rev. B {\bf 41}, 7910 (1990); L.W. Engel, S.W. Hwang, T. Sajoto, D.C. Tsui, and M. Shayegan,
Phys. Rev. B {\bf 45}, 3418 (1992).


\bibitem{footnote5aa}
We assume that both QPCs are either in a weak tunneling regime or in a weak backscattering regime.
Moreover, we further assume that two channels at each QH edge originate from the same Ohmic contact,
and therefore, they are equally biased.

\bibitem{CS-theory}
M. Marino, {\it Chern-Simons Theory, Matrix Models,
and Topological Strings} (Oxford University Press, Oxford, 2005).

\bibitem{wline}
Expression (\ref{wils}) implies that, strictly speaking,
the amplitudes $t_\ell$ in tunneling Hamiltonian (\ref{ham-0})
contain Wilson lines that connect tunneling point at opposite edges
in the left and right QPC. They have to be taken into account when evaluating
the overall AB phase shift in the tunneling amplitudes (\ref{phas-ham}).

\bibitem{klein-true}
C. de C. Chamon, E. Fradkin, Phys. Rev. B {\bf 56}, 2012 (1997).

\bibitem{Stern}
A. Stern, Ann. Phys. {\bf 323}, 204 (2008).

\bibitem{Feldman2}
D. E. Feldman and A. Kitaev, Phys. Rev. Lett. {\bf 97}, 186803 (2006); D. E. Feldman, Y. Gefen,
A. Kitaev, K. T. Law, and A. Stern, Phys. Rev. B {\bf 76}, 085333 (2007).

\bibitem{Thouless}
D.J. Thouless and Y. Gefen, Phys. Rev. Lett. {\bf 66}, 806, (1991).

\bibitem{exact}
J. Fr\"{o}hlich, B. Pedrini, in {\it Statistical Field Theories} (Kluwer Academic Publishers, Dordrecht, 2002).

\bibitem{footnote-x1}
Topological defects (vortices) of the $b$-field, if present in the bulk, change
the relation between fields $b_{\mu i}$ and $A_\mu$. Note, however, that these defects
cost a large energy (see also the discussion in Appendix \ref{choice}). Therefore,
we do not consider them.

\bibitem{footnote-x2}
It is important that at the edge $\partial_\mu j^\mu_i\neq0$, therefore the charge density
is not determined by the $b$-filed. In other words the edge deformation does not generate the
topological reconstruction in the bulk and leads only to the accumulation of the charge.

\bibitem{fr-inv}
J. Fr\"{o}hlich and E. Thiran, J. Stat. Phys. {\bf 76}, 209 (1994).

\bibitem{micro}
A.Boyarsky, V.V. Cheianov, O. Ruchayskiy, Phys. Rev. B {\bf 70}, 235309 (2004).

\bibitem{number-w}
X. Zhu, Steven G. Louie, Phys. Rev. B {\bf 52}, 5863 (1995).

\end{thebibliography}
\end{document}